\documentclass[useAMS, usenatbib,usegraphicx]{mn2e}
\usepackage{rotating}
\usepackage{amsmath}
\usepackage{url}

\newcommand{\cmt}{{\rm cm}$^{-3}$}

\newcommand{\Zsun}{Z$_\odot$}
\newcommand{\msunpc}{M$_\odot$ pc$^{-2}$}

\newcommand{\sigmol}{$\Sigma_{\rm mol}$}
\newcommand{\sigsfr}{$\Sigma_{\rm SFR}$}
\newcommand{\Lir}{$L_{\rm IR}$}
\newcommand{\Sir}{$S_{\rm IR}$}

\newcommand{\Ico}{$I_{\rm CO}$}
\newcommand{\Td}{$T_{\rm eff, i}$}
\newcommand{\Sigd}{$\Sigma_{\rm d, i}$}

\newcommand{\betaeff}{$\beta_{\rm eff, i}$}
\newcommand{\rhoTB}{$\rho_{T, \beta}$}

\newcommand{\Icounits}{K km s$^{-1}$}

\newcommand {\apgt} {\ {\raise-.5ex\hbox{$\buildrel>\over\sim$}}\ }
\newcommand {\aplt} {\ {\raise-.5ex\hbox{$\buildrel<\over\sim$}}\ }

\newcommand{\chisq}{$\chi^2$}

\newcommand{\tsig}{$2\sigma$}

\newcommand{\mICO}{I_{\rm CO}}
\newcommand{\mICOi}{I_{{\rm CO},i}}

\newcommand{\DCOi}{\mathcal{D}_{{\rm CO},i}}
\newcommand{\hmICOi}{\hat I_{{\rm CO}, i}}
\newcommand{\ICOsig}{\sigma_{\rm CO, obs}}

\newcommand{\DIRij}{\mathcal{D}_{{\rm IR}, i j}}
\newcommand{\SIRij}{S_{{\rm IR},ij}}
\newcommand{\hSIRij}{\hat S_{{\rm IR},ij}}
\newcommand{\SIRsig}{\sigma_{j}}

\newcommand{\Sfid}{S_{\rm fid}}
\newcommand{\Ifid}{I_{\rm fid}}

\newcommand{\SIRi}{S_{{\rm IR},i}}

\newcommand{\Teffi}{T_{{\rm eff},i}}

\newcommand{\beffi}{\beta_{{\rm eff},i}}

\newcommand{\likeCO}{\ell}
\newcommand{\likeIR}{k}

\newcommand{\prob}{\mathcal{P}}
\newcommand{\data}{\mathcal{D}}

\title[Modelling Dust and Gas] {Simultaneously modelling far-infrared
  dust emission and its relation to CO emission in star forming
  galaxies}
\author[R. Shetty et
al.]{Rahul Shetty$^{1}$, Julia Roman-Duval$^{2}$, Sacha Hony$^{1}$,
  Diane Cormier$^{1}$, Ralf S. Klessen$^{1}$, \and Lukas
  K. Konstandin$^{3}$, Thomas Loredo$^{4}$, Eric W. Pellegrini$^{1}$, David Ruppert$^{5}$
  \\
  $^{1}$ Zentrum f\"ur Astronomie der Universit\"at Heidelberg,
  Institut f\"ur Theoretische Astrophysik, Albert-Ueberle-Str. 2,
  69120
  Heidelberg, Germany \\
  $^{2}$ Space Telescope Science Institute, 3700 San Martin Drive,
  Baltimore, MD 21218, USA \\
  $^{3}$ School of Physics and Astronomy, University of Exeter,
  Stocker Road, Exeter EX4 4QL, UK \\
  $^{4}$ Center for Radiophysics and Space Research, Space Sciences
  Building, Cornell University Ithaca, NY 14853-6801, USA \\
  $^{5}$ Department of Statistical Science, Comstock Hall, Cornell University Ithaca, NY 14853-6801, USA \\
}
\begin{document}

\date{Accepted 2016 April 19. Received 2016 April 19; in original form 2015 September 2}

\pagerange{\pageref{firstpage}--\pageref{lastpage}} \pubyear{2015}
\maketitle

\label{firstpage}
\begin{abstract}

  We present a method to simultaneously model the dust far-infrared
  spectral energy distribution (SED) and the total infrared $-$ carbon
  monoxide (CO) integrated intensity (\Sir$-$\Ico) relationship.  The
  modelling employs a hierarchical Bayesian (HB) technique to estimate
  the dust surface density, temperature ($T_{\rm eff}$), and spectral
  index at each pixel from the observed far-infrared (FIR) maps.
  Additionally, given the corresponding CO map, the method
  simultaneously estimates the slope and intercept between the FIR and
  CO intensities, which are global properties of the observed source.
  The model accounts for correlated and uncorrelated uncertainties,
  such as those present in {\it Herschel} observations.  Using
  synthetic datasets, we demonstrate the accuracy of the HB method,
  and contrast the results with common non-hierarchical fitting
  methods.  As an initial application, we model the dust and gas on
  100 pc scales in the Magellanic Clouds from {\it Herschel} FIR and
  {\it NANTEN} CO observations.  The slopes of the log\Sir$-$log\Ico\
  relationship are similar in both galaxies, falling in the range
  1.1$-$1.7.  However, in the SMC the intercept is nearly 3 times
  higher, which can be explained by its lower metallicity than the
  LMC, resulting in a larger \Sir\ per unit \Ico.  The HB modelling
  evidences an increase in $T_{\rm eff}$ in regions with the highest
  \Ico\ in the LMC.  This may be due to enhanced dust heating in the
  densest molecular regions from young stars.  Such simultaneous dust
  and gas modelling may reveal variations in the properties of the ISM
  and its association with other galactic characteristics, such as star formation rates and/or metallicities.\\

\end{abstract}


\section{Introduction} \label{introsec}

Understanding the formation of planets, stars, and the dynamics of the
host galaxies, including galaxy clusters, invariably requires a
thorough assessment of the physical conditions of the interstellar
medium (ISM).  Targeted surveys employing ground and space based
telescopes have provided a wealth of multi-wavelength data, enabling
the concurrent study of various facets of the ISM.  For instance,
infrared (IR) and sub-millimeter observations from {\it Spitzer}
\citep{Werner+04}, {\it Herschel} \citep{Pilbratt+10}, {\it
  CARMA}\footnote{https://www.mmarray.org}, {\it
  NANTEN}\footnote{https://www.astro.uni-koeln.de/nanten2} (and other
ground based observatories), have revealed the properties of dust and
gas prevalent in the ISM, such as the temperature, chemical
composition, and density.  Flexible statistical methods and well
tested theoretical models are necessary to accurately estimate such
properties, as well as identify unanticipated features in the large
and diverse observational datasets.

The spectral energy distribution (SED) of dust can reveal its physical
characteristics.  Dust grains absorb stellar radiation, and releases
this heat in the form of far infrared (FIR) emission.  Observed FIR
intensities appear to follow a power-law modified blackbody
\citep[][]{Hildebrand83}.  Dust properties such as the temperature and
emissivity control the shape of the emergent FIR SED.  Therefore,
accurately constraining the SED parameters from IR observations
provides information about the physical characteristics of dust.
However, modelling the SED is not trivial, as there are significant
degeneracies between the parameters.  Notably, when fitting SEDs
without careful consideration of noise and the underlying correlation
between physical properties, the degeneracy between the dust
temperature and spectral index leads to an artificial anti-correlation
between the estimated parameters \citep[e.g.][]{Blain+03, Schnee+07,
  Shetty+09b, JuvelaYsard12}. Hierarchical statistical methods can
rigorously account for degeneracies and measurement uncertainties,
thereby providing accurate SED parameter estimates
\citep[e.g.][]{Kelly+12}.

On large spatial scales (\apgt 50 $-$ 100 pc), FIR emission traces
warm dust heated by the stars.  Emission from colder dust (with
temperatures \aplt 20 K) will pale in comparison, as the FIR intensity
rises strongly with temperature.  All inferred dust parameters, such
as the temperatures and column densities, must be consistent with
other constraints, including the relationship between the dust and
gas, and/or the nature of stellar radiation.  Hierarchical models are
well-suited for such analyses, as they naturally allow for the
simultaneous parameter estimates of diverse ISM components from
multi-wavelength datasets.  Such simultaneous modelling could
potentially further reveal the association, or lack thereof, of the
dust and the stellar component.

The molecular ISM is considered to be the direct precursor to the
formation of stars, since most young stars are predicted and observed
to be embedded in molecular gas \citep[see][and references
therein]{MacLow&Klessen04, McKee&Ostriker77, Fukui&Kawamura10}.  The
lowest rotational transitions of carbon monoxide (CO) are frequently
utilized as tracers of molecular gas.  In particular, the $J= 1 - 0$
transition is easily excited at typical densities ($n\approx 100$
\cmt) and temperatures (10 $-$ 100 K) of molecular clouds, and emits
at frequencies ($\approx$ 115 GHz) easily detected with ground based
sub-mm telescopes.  It is therefore one of the most widely employed
tracers of the star-forming ISM.

Given the formation of stars in the molecular ISM traced by CO, and
dust heating from young stars, there is an expectation for some
correlation between the CO and FIR intensities.  \citet{Schmidt59}
predicted that the rate of star formation (\sigsfr) should be governed
by the amount of gas through a power-law scaling.  \citet{Kennicutt89,
  Kennicutt98} indeed found a tight relationship between \sigsfr\ and
the molecular gas surface density (\sigmol) when integrating over
whole galaxies.  More recently, resolved observations have also
revealed an increasing trend of \sigsfr\ with \sigmol\
\citep[e.g.][]{Bigiel+08, Leroy+12}.  However, the indices of the
power-law relationship appear to vary between normal galaxies
\citep[e.g.][]{Bigiel+08, Shetty+13}, with most galaxies appearing to
favor a sub-linear relationship \citep[][]{Shetty+13, Shetty+14a}.
Many of these studies rely on either on monochromatic IR$-$\sigsfr\
tracers, such as 24 \micron, or a second tracer to account for the
un-absorbed stellar radiation, such as UV or H$\alpha$.  In normal and
starbursting galaxies, dust absorbs nearly all UV radiation, and so
its total IR emission is employed as a proxy for the star formation
rate, though there can be significant uncertainties in such
conversions \citep[e.g.][]{Dale+05, Pope+06}.  Dust radiative heating
from young stars is expected to decrease in low metallicity systems
\citep{Calzetti+07}, requiring alternative conversion factors compared
to normal or starbursting galaxies \citep[see review
by][]{Kennicutt&Evans12}.  Indeed, questions remain on the
relationship between the CO brightness and any star formation tracer
such as the dust luminosity, including the possible effects of other
galaxy properties, such as Hubble type, metallicity, and/or stellar
mass.

Accurately estimating any relationship between the emission from gas
and dust requires sound statistical methods.  For example, linear
regression is commonly employed for estimating the slope of the gas
$-$ SFR relationship.  Some linear regression techniques are known to
produce inaccurate parameter estimates.  For instance, when
measurement uncertainties in the predictor is ignored, the best-fit
slope will be biased towards zero \citep{Akritas&Bershady96}.
Additionally, when the dataset consists of repeated measures from a
number of individuals, fitting a single line to the pooled data
conceals variations in the parameters between individual members
within the population.  Hierarchical statistical methods can naturally
account for both of these issues, and have been shown to provide
accurate parameter estimates for both linear and non-linear models
including measurement uncertainties \citep[e.g.][]{Carrol+06,
  Gelman+04, Kelly07}.

In this work, we develop a hierarchical Bayesian method to assess the
relationship between the CO and total FIR intensities.  The method
simultaneously estimates the parameters of the dust SED at each
position, as well as the global underlying CO $-$ total FIR
relationship.  Such a hierarchical method estimates the spatial
variation in dust properties, while self-consistently measuring the
large scale relationship between dust and gas.  Consequently, the
resulting parameter estimates and their distributions probe for any
inconsistencies between the observational data and the model SED and
CO $-$ FIR relationship.  Furthermore, measurement uncertainties are
naturally propagated throughout the analysis, leading to final
parameter estimates that robustly accounts for observational noise.

As a first application, we apply the method to {\it Herschel} FIR and
{\it NANTEN} CO observations of the Large and Small Magellanic Clouds
(LMC and SMC).  Due to their proximity (50 - 60 kpc), these galaxies
allow for detailed studies of their ISM.  The Magellanic Clouds have
metallicities that are lower than the Milky Way, 0.2 \Zsun\ for the
SMC and 0.5 \Zsun\ in the LMC \citep{Russell&Dopita92}.  The factor of
$\sim$2 variation in metallicities between the Magellanic Clouds may
cause detectable differences in their dust and gas properties.  From
the HERITAGE survey data \citep[][]{Meixner+13},
\citet{Roman-Duval+14} found significant differences in the
dust-to-gas ratios between the two galaxies.  Given their large
angular size, the Magellanic Clouds present the opportunity to test
the effect of averaging over large regions, and compare any derived
trends with the small scale properties of the ISM.  Here, as an
initial application of the hierarchical Bayesian method, we model the
SED of the Magellanic Clouds on large 100 pc scales with single
modified blackbodies, in conjunction with CO maps to quantify the CO
$-$ FIR relationship.

This paper is organized as follows.  In the next section, we present
the equations governing the assumed CO $-$ FIR relationship, and the
model dust SED.  We also provide a description of hierarchical
Bayesian methods before displaying the full hierarchical model.  In
Section \ref{testsec} we demonstrate the accuracy of the model on two
synthetic datasets.  For comparison, we also present results using
common non-hierarchical fitting methods.  In the subsequent section,
we apply the hierarchical Bayesian method to observational data of the
Magellanic Clouds.  We interpret and discuss the results in the
context of previous results in Section \ref{discsec}, and summarize
the method and our findings in Section \ref{sumsec}.

\section{Modelling Method} \label{methosec}

We explore the relationship between gas and dust throughout a galaxy
by characterizing the relationship between CO and FIR emission across
the projected face of the galaxy, treating the lines of sight
corresponding to pixels as probing regions that independently sample a
global, stochastic CO--FIR relationship.  A more sophisticated
treatment would additionally model spatial correlation and dependence.
Our focus here is to develop and implement a framework that can
correctly account for pixel-specific variability (from measurement
error).  Our hierarchical framework can be generalized to account for
spatial dependence, but we leave that for future work.

\subsection{Underlying relationships}

Thermal emission from dust grains is usually modelled with a power-law
modified Plank spectrum.  The observed surface brightness $S_\nu$ at a given
frequency $\nu$ is:
\begin{equation}
S_{\rm \nu} = \Sigma_{\rm d} B_{\rm \nu}(T_{\rm eff}) \kappa_0
(\nu/\nu_0)^{\beta_{\rm eff}}
\label{Snueqn}
\end{equation}
where $\Sigma_{\rm d}$ is the dust surface density\footnote{In cgs
  units, $\Sigma_{\rm d}$ as written in equation \eqref{Snueqn}
  corresponds to g cm$^{-2}$.  In this work we convert
  $\Sigma_{\rm d}$ to \msunpc.}, $T_{\rm eff}$ is the dust
temperature, and $\kappa_\nu=\kappa_0 (\nu/\nu_0)^{\beta_{\rm eff}}$
is the frequency dependent dust opacity, which depends on the spectral
index $\beta_{\rm eff}$.  The SED of a pure blackbody follows the
Planck function:
\begin{equation}
B_{\rm \nu}(T_{\rm eff}) = \frac{2 h \nu^3/c^2}{\exp(h\nu/k_BT_{\rm eff})-1}
\label{planck}
\end{equation}
where $h$, $c$, and $k_B$ are the Planck constant, speed of light, and
Boltzmann constant, respectively.

Equation \eqref{Snueqn} is a simplified model of any emergent SED, as
it employs a number of approximations about the dust along the line of
sight (LoS).  As dust absorbs the radiation from young stars, the dust
temperature depends on the distances to the these stars.
Consequently, a single $T_{\rm eff}$ does not accurately model the
emergent SED.  The estimated temperature may better reflect a
luminosity weighted temperature, and is an upper limit to the coldest
temperature along the LoS \citep[e.g.][]{Shetty+09b, Malinen+11,
  JuvelaYsard12}.  Similarly, dust grains along the LoS can vary in
composition or size and may have a range of spectral indices.
Accordingly, we will only consider the estimated temperature and
spectral index to be adequate approximations for describing the shape
of the emergent SED.  Following the convention of \citet{Gordon+14},
we will therefore refer to these quantities as the ``effective''
temperature or spectral index (hence the subscript on $T_{\rm eff}$
and $\beta_{\rm eff}$).

The total FIR intensity \Sir\ can be computed by integrating the SED over
all frequencies:
\begin{equation}
S_{\rm IR} = \int S_{\rm \nu} d\nu = \Sigma_{\rm d} \int  B_{\rm \nu}(T_{\rm eff}) \kappa_\nu d\nu
\label{IRint}
\end{equation}
We model the relationship between the CO intensity \Ico\ and \Sir\
through a power-law, which translates to a linear trend in log-space:
\begin{equation}
\log (S_{\rm IR}/\Sfid) = A + n \log
(I_{\rm CO}/\Ifid)
\label{Lcoir}
\end{equation}
where $\Sfid=1 \, {\rm MJy \, Hz \, sr}^{-1}$ and
$\Ifid=1 \, {\rm K \, km \, s}^{-1}$ are fiducial values to make the
arguments of the logarithms dimensionless.\footnote{In subsequent
  equations, we will omit these fiducial quantities in the logarithms
  for brevity and to follow convention.}  Note that when CO is assumed
to be a linear tracer of molecular gas, \Ico\ will be proportional to
\sigmol.  A common assumption for normal star-forming galaxies is that
dust is mostly heated by newly born stars, so that dust thermal
emission indirectly traces the amount of star formation.\footnote{As
  we further discuss in Section \ref{discsec}, other properties
  besides SFR influence the dust surface density and temperature that
  affect the emergent SED.}  Since the amount of dust heating depends
on metallicity, among other ISM properties, the FIR $-$ SFR
relationship may vary with environment.  We choose not to employ any
conversion factor, and only focus on estimating the total FIR $-$ CO
relationship.  Note that if there were any constant FIR $-$ SFR
scaling, the slope in equation \eqref{Lcoir} is the exponent in the
Kennicutt-Schmidt (KS) relationship:
\begin{equation}
\Sigma_{\rm SFR} = a \Sigma_{\rm mol}^n.
\label{KSlaw}
\end{equation}

For given values of $A$ and $n$, the observed CO map and equation
\eqref{Lcoir} set \Sir\ at each location.  Combined with the dust SED
model in equations \eqref{Snueqn} $-$ \eqref{IRint}, the dust surface
density is constrained:
\begin{eqnarray}
\log \Sigma_{\rm d} = \log S_{\rm IR} - \log \left( \int B_{\rm \nu}(T_{\rm eff}) \kappa_\nu d\nu \right ) \\
\log \Sigma_{\rm d} = A + n\log I_{\rm CO} - \log \left( \int B_{\rm \nu}(T_{\rm eff}) \kappa_\nu d\nu \right ).
\label{Ndeqn}
\end{eqnarray}
Equation \eqref{Ndeqn} relates the dust surface density $\Sigma_{\rm d}$
with the true CO intensity \Ico, and the integrated SED
$\int B_{\rm \nu}(T_{\rm eff}) \kappa_\nu d\nu$.

\subsection{Measurement Uncertainties} \label{mmsec}

Let $\DCOi$ denote the data used to estimate $\mICOi$, the CO
intensity for location $i$.  A data processing pipeline produces a
measured intensity for the location, $\hmICOi(\DCOi)$, and an
uncertainty for the intensity, $\ICOsig(\DCOi)$.  We interpret these
as summaries of a likelihood function for the CO intensity at location
$i$ that is log-normal, i.e., Gaussian in $\log(\mICO)$ (at least to a
good approximation).%
\footnote{Formally, this is most likely a marginal or profile
  likelihood function, in that modeling the data will require
  estimating parameters in addition to the CO intensity, such as
  background parameters.  Uncertainty in these parameters may be
  propagated into the CO intensity estimate by marginalization
  (integration) or profiling (optimization).}  We denote this
likelihood function by
\begin{align}
\likeCO_i(\mICOi)
  &\equiv p(\DCOi|\mICOi)\nonumber\\
  &\propto \exp\left[-\frac{1}{2\ICOsig^2} \left(\log\frac{\hmICOi}{\mICOi}\right)^2\right].
\label{like-CO}
\end{align}

Similarly, let $\DIRij$ denote the data used to estimate $\SIRij$, the
IR intensity for location $i$ in frequency channel $j$.  We denote the
associated measured intensity by $\hSIRij(\DIRij)$, and the intensity
uncertainty by $\SIRsig(\DIRij)$, which we consider to be independent
between pixels.  Besides the usual random uncertainties (independent
across pixels and channels), \emph{Herschel} intensity estimates also
have systematic uncertainty due to absolute calibration uncertainty
(see \S~4.1).  To account for this, we include channel-dependent
calibration parameters corresponding to an uncertain multiplicative
factor in intensity, and thus additive in logarithm of intensity; we
denote the additive parameter by $C_j$.  The likelihood function for
$\SIRij$ and $C_j$ is
\begin{align}
\likeIR_{ij}(\SIRij, C_j)
  &\equiv p(\DIRij|\SIRij, C_j)\nonumber\\
  &\propto \exp\left[-\frac{1}{2\SIRsig^2}\left(\log\frac{\hSIRij}{C_j\SIRij}\right)^2\right].
\label{like-IR}
\end{align}
We take $C_j$ to be the same constant across all channels within a
particular instrument (i.e., PACS and SPIRE), so that $C_1 = C_2
\equiv C_{\rm PACS}$, and $C_3 = C_4 = C_5 \equiv C_{\rm SPIRE}$.

We note that in our hierarchical modelling, we approximate all
uncertainties with log-normal distributions.  This choice is motivated
by the convenience of transforming all intensities into log space.
Since the normally distributed uncertainties are all of order 10\% or
less, a log-normal approximation is adequate.  We have verified with a
few simple tests that modelling normally distributed errors in the CO
and IR intensities as log-normals in the HB model accurate recovers
the underlying latent parameters in the posterior.

%-------------------------------------------------------------------------------
\subsection{The Hierarchical Model} \label{hiermodsec}

We adopt a Bayesian approach, addressing parameter estimation questions by computing the posterior probability density function (PDF) for parameters $\Theta$ given the observed data $\data$, denoted $\prob(\Theta|\data)$.
Bayes's theorem expresses the posterior PDF in terms of more accessible PDFs:
\begin{equation}
\mathcal{P}(\Theta|\data)
 = \frac{\prob(\Theta) \prob(\data|\Theta)}{\prob(\data)}
 = \frac{\prob(\data,\Theta)}{\prob(\data)}.
\label{BayesThm}
\end{equation}
That is, the posterior is proportional to the product of a prior PDF, $\prob(\Theta)$, and a likelihood function $\prob(\data|\Theta)$, which is the probability of observing the data $\data$ given $\Theta$ (considered as a function of $\Theta$).
Equivalently, the posterior is proportional to the joint distribution for the data and parameters, $\mathcal{P}({\mathcal{D},\Theta})$.
The term in the denominator, $\mathcal{P}(\mathcal{D})$, is constant with respect to the parameters, playing the role of a normalization constant.
It is the prior predictive distribution for the data, also called the marginal likelihood.

Since the posterior provides a probability density for a set of parameters (conditional on the data), in the Bayesian framework the estimated parameters are considered to be random variables themselves---but ``random'' in the sense of \emph{uncertain}, rather than in the frequentist sense of varying upon repetition of the experiment.

We will build up to our full hierarchical model for the CO and IR data across a galaxy image by considering three simpler inference problems that will appear as components of our full model.

\begin{figure}
\includegraphics*[width=90mm]{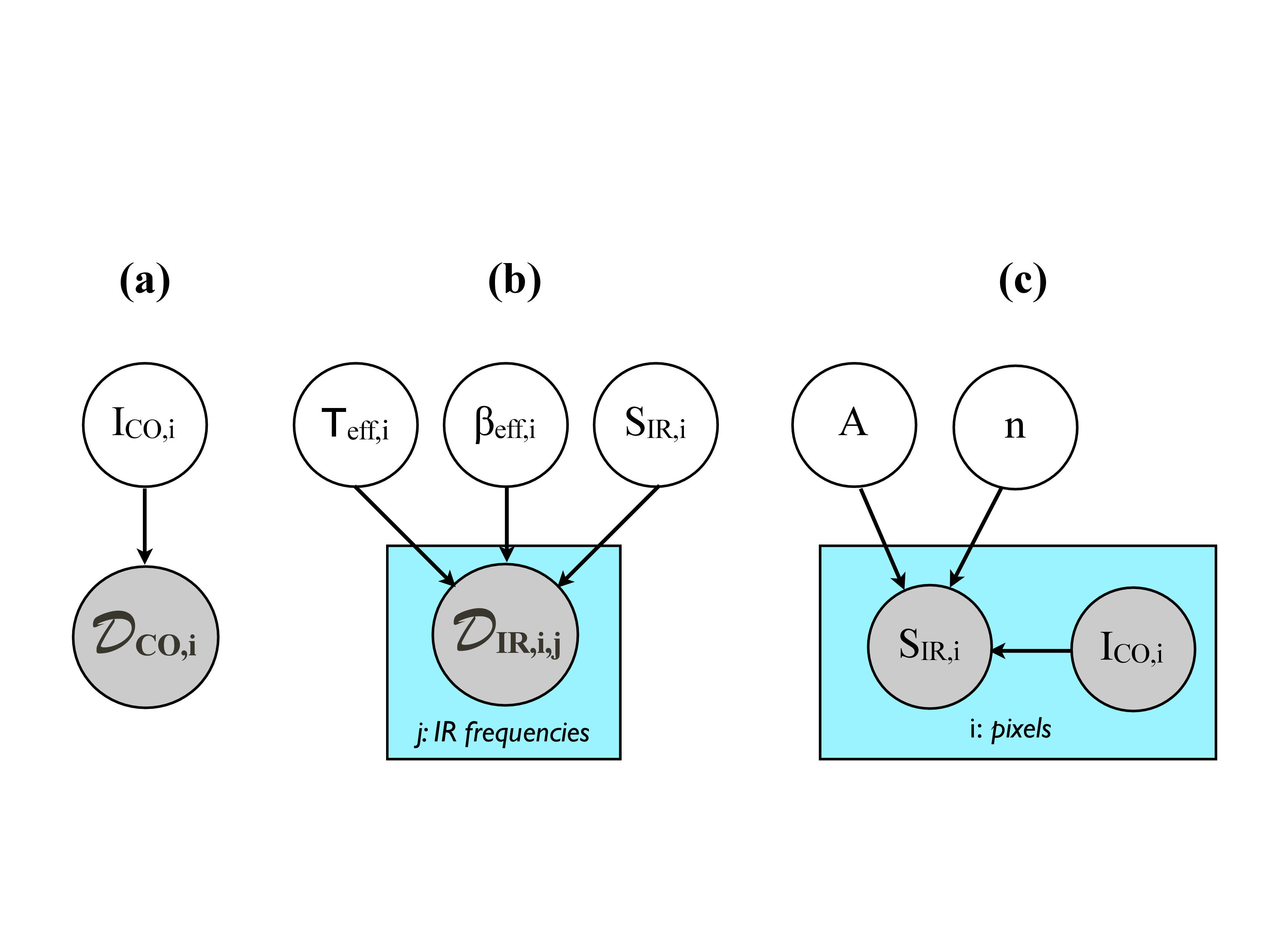}
\caption{Example Directed Acyclic Graphs (DAG) showing conditional
  dependencies between parameters (white circles) and measured data
  (gray circles). (a) DAG indicating that the measured CO data
  $\mathcal{D_{\rm CO, i}}$ can be predicted given the true CO
  intensity $\mICOi$.  (b) DAG where \Td, \betaeff, and $\SIRi$
  determine the IR data $\mathcal{D_{\rm IR,i,j}}$ at each frequency
  $j$.  (c) DAG where the data are simply the IR and CO intensities,
  which are conditionally dependent on the regression parameters $A$
  and $n$.  See Section \ref{hiermodsec}.}
\label{fig:DAGs}
\end{figure}

First, suppose we are solely interested in the CO intensity for a
single location, so the parameter space is $\Theta = \mICOi$.  We
might adopt a flat prior for $\mICOi$, in which case the posterior PDF
for $\mICO$ would be proportional to the CO pixel likelihood factor of
equation \eqref{like-CO} (for the pixel of interest).
Figure~\ref{fig:DAGs}(a) depicts the conditional structure of this
elementary application of Bayes's theorem with a directed acyclic
graph (DAG), with nodes corresponding to random (a priori uncertain)
quantities (parameters or data), and directed edges (arrows)
indicating conditional dependence.  The single edge here simply
indicates that the modelling information lets us predict the data when
the parameter, $\mICO$, is known.  The data node is shaded gray to
denote that the data become known at the time of analysis.  As a
whole, the DAG describes a factorization of the joint distribution,
$p(\mICO,\mathcal{D}_{\rm CO}) = p(\mICO) p(\mathcal{D}_{\rm
  CO}|\mICO)$, i.e., the numerator in Bayes's theorem.

Next, suppose we are solely interested in the dust spectrum
parameters, $\Theta = (\SIRi, \Teffi, \beffi)$, for a particular
location $i$, given the IR data for all channels.  The likelihood
function for the spectral parameters is the product of IR likelihood
factors given by equation \eqref{like-IR}, with the $\SIRij$ values
computed using the spectrum parameters.  For simplicity, we ignore the
intensity calibration parameters for the moment.  Then the likelihood
function for the spectrum parameters is
\begin{align}
p(\mathcal{D}|\Theta)
  &= \prod_j \likeIR_j[\SIRij(\Theta)] \nonumber\\
  &\propto \exp\left[-\frac{\chi_{{\rm IR},i}^2(\Theta)}{2}\right],
\label{locnlike-IR}
\end{align}
where $\chi_{{\rm IR},i}^2(\Theta)$ is the familiar goodness-of-fit measure,
\begin{equation}
\chi_{{\rm IR},i}^2(\Theta)
  = \sum_j \frac{[\log(\hSIRij/\Sfid) - \log(\SIRij(\Theta)/\Sfid)]^2}{\sigma_{j}^2}.
\label{chisqr-IR-px}
\end{equation}
If a flat prior PDF is adopted for the spectrum parameters, the mode (the parameter vector that maximizes the posterior PDF) is the maximum likelihood estimate, which is the parameter vector that minimizes $\chi_{{\rm IR},i}^2(\Theta)$.

Figure~\ref{fig:DAGs}(b) depicts the conditional structure of this use
of Bayes's theorem.  The top nodes denote independent prior PDFs for
the three parameters.  The three arrows indicate that all three
parameter values must be specified to predict the data.  The five data
nodes (for the five frequency channels) are depicted using a
\emph{plate}, a box with a label indicating repetition over a
specified index (here the channel index, $j$).  Crucially, a plate
signifies that each case is conditionally independent of the others;
that is, with $\Theta$ given, the probability for $\mathcal{D}_2$
(say) is independent of the values of the data in other channels.

Finally, suppose that the (true) CO intensity, $\mICOi$, and total IR
luminosity, $\SIRi$, were measured precisely for $N$ pixels (i.e., the
data are the precise values, rather than uncertain estimates from
photometry).  In this case, we could learn the dust-gas relationship
parameters in equation \eqref{Lcoir}, $\Theta = (A,n)$, via
regression, i.e., using the $N$ pairs $(\mICOi, \SIRi)$, to infer the
conditional expectation of $\SIRi$ given $\mICOi$ (and $\Theta$).  The
DAG in Figure~\ref{fig:DAGs}(c) shows the conditional structure of
this regression model.

Commonly, a regression analysis quantifies the scatter about the fit
line.  In the next section, we compare the results from the
hierarchical modelling with a non-hierarchical approach which utilizes
standard regression methods.  If we suppose the scatter about the
regression line is Gaussian with standard deviation $s$, the
likelihood function for estimating $\Theta = (A,n)$ in this problem is
a product of normal distributions for the residuals,
\begin{align}
p(\mathcal{D}|\Theta)
  &= \prod_{i=1}^N  \frac{1}{s\sqrt{2\pi}} \exp\left\{-\frac{1}{2s^2}[\log(\SIRi/\Sfid) \right.\\
  &\qquad \left.\vphantom{\frac{1}{2s^2}}- A - n\log(\mICOi/\Ifid) ]^2\right\}\nonumber\\
  &\propto \frac{1}{s^N}\exp\left[-\frac{\chi_{{\rm LR}}^2(\Theta)}{2}\right],
\end{align}
where $\chi_{{\rm LR}}^2(\Theta)$ is the sum of squared residuals
(normalized by $s^2$) that is minimized in least-squares linear
regression (LR),
\begin{equation}
\chi_{{\rm LR}}^2(\Theta)
  = \sum_i \frac{[\log(\SIRi/\Sfid) - A - n\log(\mICOi/\Ifid)]^2}{s^2},
\label{chisqr-LR}
\end{equation}

To analyze the CO-FIR data, we are ultimately interested in regression
and estimation of $(A,n)$.  However, we do not have precise
$(\mICOi, \SIRi)$ pairs; we have CO and IR data, providing uncertain
estimates of these quantities.  Perhaps the simplest way forward is to
ignore the $\mICOi$ and $\SIRi$ uncertainties (or hope they will
``average out''), using the estimates as if they were precise values
in a linear regression model like that depicted in
Figure~\ref{fig:DAGs}(c).  But problems of this sort are well-studied
in the statistics literature on \emph{measurement error} and
\emph{errors-in-variables} problems, where it is known that, instead
of averaging out, the uncertainties instead accumulate, producing
\emph{inconsistent} parameter estimates (i.e., estimates that converge
to incorrect values as data accumulate).  Multilevel or hierarchical
models provide a flexible framework for full accounting of such
uncertainties, avoiding these inconsistencies.

\begin{figure}
\includegraphics*[width=90mm]{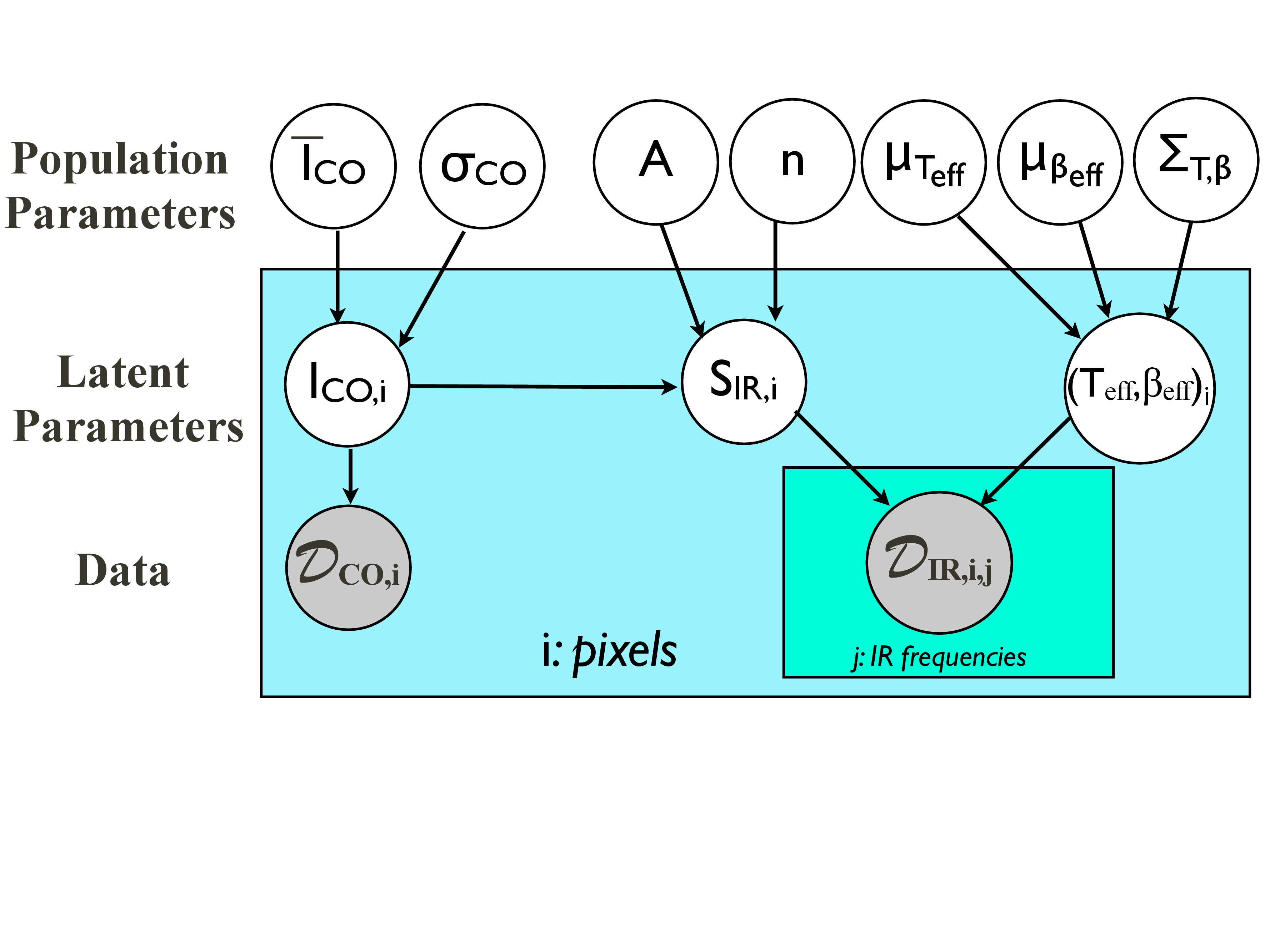}
\caption{DAG showing the conditional dependencies between the
  estimated parameters and data of the hierarchical model described in
  Section \ref{methosec}.  The highest level contains the global
  population, or hyper-, parameters such as the mean dust temperature
  ($\mu_T$), the \Td\ and \betaeff\ covariance matrix
  $\Sigma_{T, \beta}$, which includes their correlation
  ($\rho_{T,\beta}$), or the Kennicutt-Schmidt index ($n$).  These
  parameters are required to estimate the other latent parameters
  occurring at the pixel level, denoted with a subscript $i$, and are
  located within the cyan box, or plate, in the DAG.  The subscript
  $j$ refers to the frequency of the observed IR intensity, located
  within an additional plate.  The dust surface density \Sigd\ can be
  evaluated given the other parameters (via eqn. \ref{Nks}), and is
  required to estimate the IR intensity $S_{i,j}$.}
\label{dag}
\end{figure}

Here we build a hierarchical model by composing the three DAGs
described above (but leaving the $\mICOi$ and $\SIRi$ nodes open in
the regression DAG, since the true values of these quantities are
uncertain).  In our HB model for the CO and FIR data, the parameter
space is large (and grows in size with the data); $\Theta$ includes
the regression parameters, $(A,n)$, the uncertain CO and IR
intensities, $(\mICOi, \SIRi)$, and spectral parameters for each
pixel, $(\Teffi,\beffi,\SIRi)$.  Figure~\ref{dag} shows the DAG
connecting all of these quantities; the DAGs of Figure~\ref{fig:DAGs}
appear as sub-structures within this HB model.  Note that for the
first two DAGs in Figure~\ref{fig:DAGs}, we were focusing our
attention on a single location (pixel); in the single-location
analyses described above, we specified flat prior PDFs for $\mICOi$
and the IR SED parameters (in the absence of better-motivated
alternatives).  In the HB model, we jointly analyze data from all
locations.  This enables us to learn about the population
distributions for these parameters.  We do so by parameterizing these
distributions, e.g., adopting a log-normal distribution for $\mICOi$
and a bivariate normal for $(\Teffi,\beffi)$ (details are provided
below).  The uncertain parameters specifying these population
distributions become additional nodes in the HB model, as shown in
Figure~\ref{dag}; in HB terminology, these are \emph{hyperparameters}.
For more information on hierarchical Bayesian methods, we refer the
reader to \citet{Gelman+04}, \citet{Gelman&Hill07}, and
\citet{Kruschke11}.

%=========== Only relatively minor mods below =============

A DAG only specifies the qualitative structure of the HB model; to implement it, we must specify distributions (suitably conditioned) for every node.
For instance, the DAG indicates that we need a distribution for the pair $(\Teffi,\beffi)$ conditional on their mean values and the covariance matrix ($\mu_T$, $\mu_{\beta}$, $\Sigma_{T, \beta}$), independent of all other parameters.
We model the dust \Td\ and \betaeff\ with a bivariate normal PDF.
Therefore, that node of the DAG corresponds to a factor in the joint PDF:
\begin{equation}
  \mathcal{P}(T_{\rm eff},\beta_{\rm eff}|\mu_T, \mu_\beta, \Sigma_{T, \beta}) =
  \mathcal{N}(T_{\rm eff},\beta_{\rm eff}|\mu_T, \mu_\beta, \Sigma_{T,
    \beta}),
\label{TDnormexm}
\end{equation}
where $\mathcal{N}({\bf x}|{\bf S})$ is the bivariate normal PDF for ${\bf x}$ given means and covariance ${\bf S}$.
We employ simpler, abbreviated standard statistical notation, so that equation \eqref{TDnormexm} is written as
\begin{equation}
(T_{\rm eff}, \beta_{\rm eff})^T | {\bf \mu_{T,\beta},  \Sigma_{T, \beta}} \sim {\bf
  \mathcal{N}}({\bf \mu}_{T,\beta}, {\bf \Sigma_{T, \beta}}),
\label{TDnormexmII}
\end{equation}
where $({\bf x})^T$ is the transpose of the vector ${\bf x}$, and ${\bf \mu_{T,\beta}} = (\mu_T, \mu_\beta)^T$.
Some parameters, such as the correlation between \Td\ and \betaeff, \rhoTB, are modeled with uniform distributions, denoted by $\mathcal{U}(min, max)$ spanning $min$ and $max$.

In order to evaluate the joint likelihood of all the parameters, we
model the distributions of most parameters as normals.  The quantities
$A$, $n$, $\overline{I}_{\rm co}$ and ${\sigma}_{\rm co}$ are the
population parameters describing the intercept and slope of the
\Sir$-$\Ico\ relationship (Eqn.  \ref{Lcoir}), and the mean and
standard deviation of the true CO intensity (in log space),
respectively.  Obtaining a reliable estimate for these quantities is
one of the primary goals of the hierarchical fitting process.  The
distributions of the population parameters also require (hyper)
priors.  Again, we choose normal distributions with large variances.
The choice of the mean values of these hyperpriors does not affect the
posterior, again due to the large number of datapoints that constrain
these parameters more than the priors.  In our tests of the method
described in the next section, we investigate the influence of the
prior distributions when the underlying data are not normal.

For the CO and IR data nodes, we adopt log-normal likelihood functions
as described above, in equations \eqref{like-CO} and \eqref{like-IR}.
We assign a population distribution to the CO intensities that is
log-normal, specified by two hyperparameters (population distribution
parameters): a mean, $\log \overline{I}_{\rm CO}$, and standard
deviation, $\sigma_{\rm CO}$.  We assign these parameters a normal and
a uniform prior, respectively.  Thus,
\begin{align}
\log ({I}_{\rm CO,i}/\Ifid) | \log (\overline{I}_{\rm CO}/\Ifid), \sigma_{\rm CO}
  &\sim \mathcal{N}(\log \overline{I}_{\rm CO}/\Ifid, \sigma_{\rm CO}^2), \\
\log (\overline{I}_{\rm CO}/\Ifid)
  &\sim \mathcal{N}(0, 5) \label{meanCO}, \\
\sigma_{\rm CO}
  &\sim \mathcal{U}(0, 2) \label{sigco}.
\end{align}
We treat each source's temperature and power law index as drawn from a bivariate normal population distribution,
\begin{equation}
(T_{\rm eff}, \beta_{\rm eff})_i^T | {\bf \mu_{T,\beta},  \Sigma_{T, \beta}}
  \sim {\bf \mathcal{N}}({\bf \mu_{T,\beta}}, {\bf \Sigma_{T, \beta}}) \label{TB_MVN}.  \\
\end{equation}
The population mean temperature and index have normal hyperpriors;
\begin{align}
\mu_{T}
  &\sim \mathcal{N}(40, 20) \label{muT},  \\
\mu_{\beta}
  &\sim \mathcal{N}(0, 3) \label{muB}.
\end{align}
We write the temperature-index covariance matrix in terms of the (marginal) standard deviations for temperature and index, and a correlation parameter, $\rho_{T, \beta}$:
\begin{equation}
{\bf \Sigma_{T, \beta}} | \rho_{T, \beta}, \sigma_T, \sigma_\beta =
\begin{bmatrix}
\sigma_T^2 & \rho_{T, \beta}\sigma_T\sigma_\beta \\
\rho_{T, \beta}\sigma_T\sigma_\beta & \sigma_\beta^2
\end{bmatrix}.
\end{equation}
We assign uniform priors to these three hyperparameters:
\begin{align}
\rho_{T, \beta}
  &\sim \mathcal{U}(-1, 1) \label{rhoTB},  \\
\sigma_T
  &\sim \mathcal{U}(0, 50) \label{sigT},  \\
\sigma_\beta
  &\sim \mathcal{U}(0, 3) \label{sigB}.
\end{align}
Note that it is customary in non-hierarchical models to assign
unknown standard deviations log-uniform priors (i.e., priors
proportional to the inverse standard deviation).  In a hierarchical
setting, because of the weakened connection between hyperparameters
and the data, such an assignment can unduly influence the posterior
(even making it improper, i.e., unnormalizeable).  Flat priors have
better behavior while still remaining only weakly informative
\citep{Gelman+04}.

Finally, we specify informative but broad priors on the offset ($A$) and slope ($n$) in the $\log(S)$--$\log(I)$ relationship,
\begin{align}
A
  &\sim \mathcal{N}(0, 100) \label{alphadist},  \\
n
  &\sim \mathcal{TN}(0, 4) \label{ndist},
\end{align}
with $\mathcal{TN}(\cdot,\cdot)$ denoting a truncated normal,
requiring the quantity to be greater than or equal to the specified
mean (here requiring $n\ge 0$).  These priors are motivated by
previous observational studies, which always find and increasing
\sigsfr\ with \sigmol.  We follow convention and assume constant
conversions between FIR and \sigsfr, as well as CO and \sigmol.
However, we do not favor any particular values for the conversion
factors, hence employing a large range for the possible value of the
offset ($A$) parameter.

The dust surface density can be computed from the modeled parameters described above:
\begin{multline}
\log \Sigma_{\rm d, i} | A, n, I_{\rm CO,i}, T_{\rm eff, i}, \beta_{\rm eff, i}  = \\
A + n\log(I_{\rm CO,i}/\Ifid) - \log \left(\int  B_{\rm \nu}(T_{\rm eff, i})
  \kappa_0 \left (\frac{\nu}{\nu_0} \right)^{\beta_{\rm eff, i}} d\nu \right)
\label{Nks}
\end{multline}
With \Sigd, \Td\, and \betaeff, the true IR intensity at each pixel
$i$ is given by Equation \ref{Snueqn}:
\begin{multline}
{S}_{\rm i, j} | (T_{\rm eff}, \beta_{\rm eff})_i, \Sigma_{\rm d, i} =\\
  \Sigma_{\rm d, i} \frac{2 h \nu_j^3}{c^2} \kappa_0 \left
  (\frac{\nu_j}{\nu_0} \right)^{\beta_{\rm eff, i}}
\frac{1}{\exp(h\nu_j/k_BT_{\rm eff, i})-1}.
\label{modPlanck}
\end{multline}
The reference opacity at $\nu_0$=230 GHz is assumed to be fixed at
$\kappa_0=0.9$ cm$^{2}$ g$^{-1}$ \citep{OssenkopfHenning94}.

We note that we do not favor any particular \rhoTB\, so it is modeled with a uniform distribution between $-$1 (fully anti-correlated) and 1 (exact correlation).
We model the mean temperature and spectral index, $\mu_T$ and $\mu_\beta$, as normal distributions, with large variances for their priors, allowing the sampler to explore a wide range of possible values near typical ISM conditions.
Given the large number of datapoints, the final estimates for $\mu_T$ and $\mu_\beta$ are not sensitive to these choices.

To carry out the hierarchical analysis, we use the {\tt Stan}
probabilistic programming language via its R language API
\citep{rstan}.\footnote{{\tt Stan} is publicly available at
  http://mc-stan.org/index.html}$^{\rm ,}$\footnote{In {\tt Stan},
  we recover the correct likelihoods when we model the observed CO
  intensities as
  $\log (\hat{I}_{\rm CO,i}/\Ifid) | \log (I_{\rm CO,i}\Ifid),
  \sigma_{\rm CO, obs} \sim \mathcal{N}(\log (I_{\rm CO,i}\Ifid),
  \sigma_{\rm CO, obs}) $
  and the FIR with
  $\log (\hat{S}_{\rm i, j}/\Sfid) | \log ({S}_{\rm i, j}/\Sfid),
  \sigma_{\rm j} \sim \mathcal{N}(\log({S}_{\rm i, j}/\Sfid) +
  \mathcal{C_{\rm j}}, \sigma_{\rm j} )$.}
{\tt Stan} performs efficient sampling of the parameters through a
Hamiltonian Monte Carlo algorithm.  We refer the reader to the manual
and website for more information about the details of {\tt Stan}.

\section{Simulation Studies} \label{testsec}

To test the HB fitting method, we construct synthetic datasets and
compare the posterior with the adopted underlying parameters.  For
some parameters, we also investigate the effect of choosing incorrect
prior distributions, e.g. a truncated normal distribution for \Ico\
when the underlying distribution is in fact log-normal.  In this
section, we present and discuss the results from three such
investigations, though we have performed the HB analysis on several
realizations of the synthetic datasets.  Here, we aim to demonstrate
that the HB model described in Section \ref{methosec} is properly
implemented.  We also illustrate that the HB analysis provides more
accurate parameter estimates than a simple regression analysis.  For
the latter, we construct synthetic datasets for which the underlying
parameters do not follow the distributions assumed in the HB model.

The synthetic datasets are characterized by five latent variables:
$\log$ \Sigd, $\log$ \Sir, $\log$ \Ico, \Td, and \betaeff.  From these
quantities and chosen observational uncertainties, we produce the
observed CO and IR intensities, $\hat I_{\rm CO, i}$ and
$\hat{S}_{\rm i, j}$.  The first two columns of Tables
\ref{bayes_DS_A} and \ref{bayes_DS_B} show the summary statistics of
the most interesting hyperparameters of the synthetic datasets.

\begin{figure*}
\includegraphics*[width=140mm]{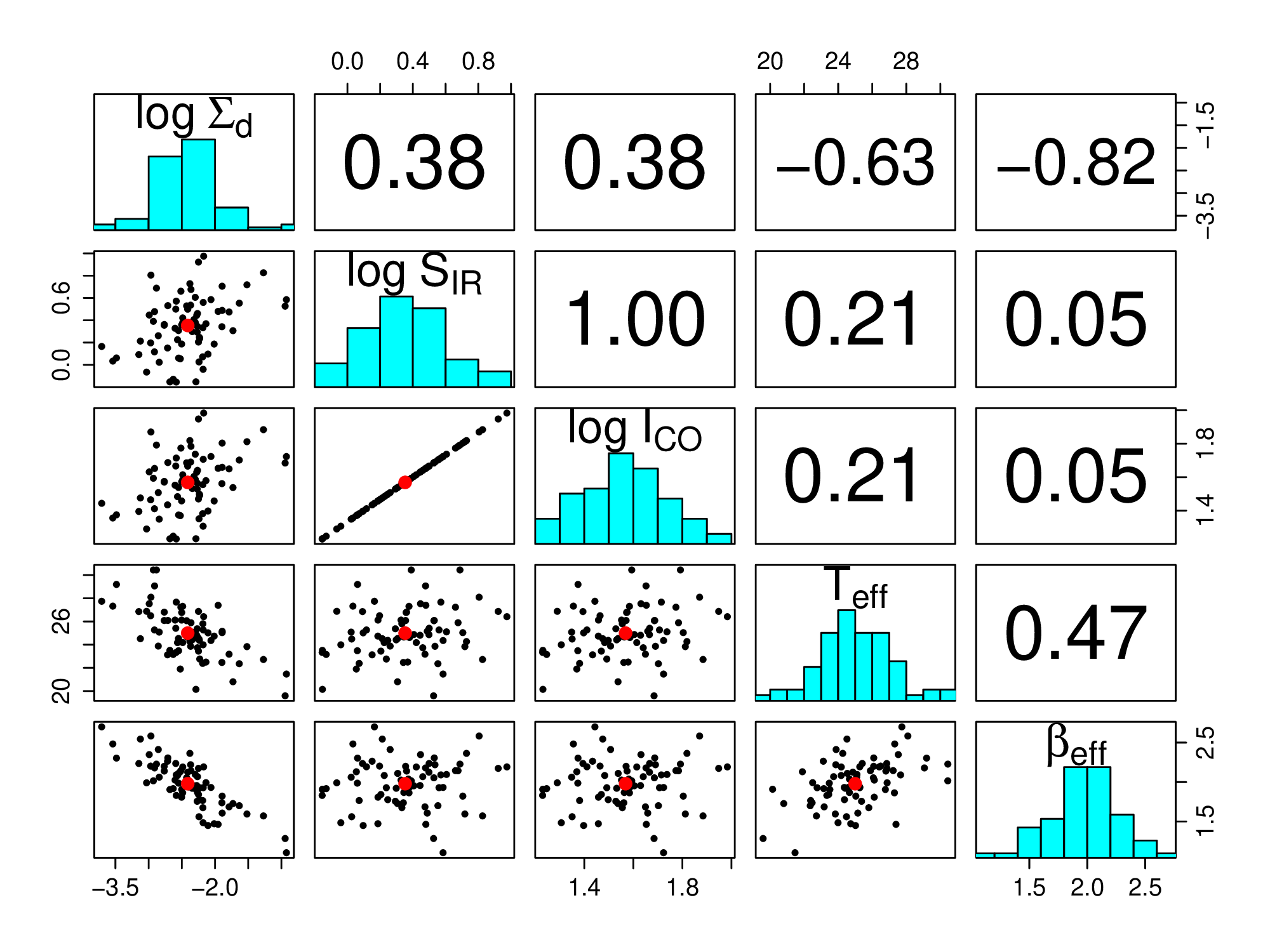}
\caption{Distributions across pixel and bivariate relationships of
  quantities in synthetic dataset A.  The diagonal panels label the
  variables associated with that row and column, and include their
  histograms.  The upper triangle lists the correlation coefficient
  between the quantities in the same row and column.  Panels in the
  lower triangle show the scatter plot of the corresponding data, with
  the red points indicating the mean values.}
\label{testA}
\end{figure*}

Figure \ref{testA} shows the distributions and bivariate relationships
for the true values of the 5 individual pixel latent variables in
synthetic dataset A.  The panels on the diagonal display the
histograms of each variable.  The panels below the diagonal show the
scatter-plots between the variables identified by its row and column
position.  Similarly, the correlation coefficient between the
corresponding variables are shown in the panels above the diagonal.
For synthetic dataset A, we choose distributions of the underlying
parameters to match those in the HB modelling, as the aim of this
initial test is verify that the HB model is implemented correctly.  In
Table \ref{bayes_DS_A}, the second column lists the adopted values of
the parameters in test dataset A.  For each of the 75 replicates (or
``pixels'') of this dataset, we include a 15\% uncertainty to create
the synthetic observed CO intensities, and a random 2\% uncertainty on
the IR intensity.  For the calibrated errors, we employ 10\% and 8\%
uncertainties for $\mathcal{C_{\rm PACS}}$, and
$\mathcal{C_{\rm SPIRE}}$, respectively, corresponding to the
estimated uncertainties from the PACS and SPIRE instruments (see
Section \ref{uncsec}).

\begin{table*}
 \centering
 \begin{minipage}{140mm}
   \caption{Adopted and HB Estimated Population Parameters for
     Synthetic Dataset A$^1$}
  \begin{tabular}{cccc}
    \hline
    \hline
    Parameter & True Value & Posterior Mean & 95\% HPD \\
    \hline
    $\rho_{T,\beta}$ & 0.47 & 0.38 & [0.18, 0.59]  \\
    $\mu_{T_{\rm eff}}$ & 25.0 & 25.1 & [24.7, 25.6]  \\
    $\sigma_T$ & 2.1 & 2.2 & [1.8, 2.6]  \\
    $\mu_{\beta_{\rm eff}}$ & 1.98 & 1.97 & [1.91, 2.05]  \\
    $\sigma_\beta$ & 0.3 & 0.32 & [0.26, 0.37]  \\
    $\log \overline{I}_{\rm CO}$ & 1.57 & 1.56 & [1.52, 1.60]  \\
    $\sigma_{\rm CO}$ & 0.17 & 0.17 & [0.14, 0.20]  \\
    $n$ & 1.5 & 1.49 & [1.37, 1.63]  \\
    $A$ & $-$2.00 & $-$1.95 & [$-$2.15, $-$1.75]  \\
    \hline
    \footnotetext[0]{$^1$ This dataset contains 75 repeated measures of CO and
    IR luminosities, including 2\% and 10\% noise uncertainties,
    respectively, along with 10\% correlated IR uncertainties.  The
    effective sample size $\approx$4000 and $\hat{R}\approx$ 1.}
\end{tabular}
\label{bayes_DS_A}
\end{minipage}
\end{table*}

In sampling the posterior, we run three MCMC chains and ensure
sufficient mixing and convergence by inspecting that the $\hat{R}$
values of all the population parameters are very close to one, and
that the effective sample size\footnote{We also inspect latent
  parameters of a few replicates (pixels), such as \Td\ and \betaeff.
  For these parameters, we also find $\hat{R} \approx 1$, with very
  high effective sample sizes (\apgt 15,000).} is large
\citep{Gelman+04, Flegal+08}.  Since we are interested in 95\% density
intervals, we ensure that the effective sample sizes of the main
population parameters of interest, $\rho_{T,\beta}$,
$\mu_{\beta_{\rm T}}$, $\mu_{\beta_{\rm eff}}$, $n$, and $A$, are at
least 3500, with corresponding $\hat{R}\approx 1$.  We choose random
initial values for the latent parameters, ensuring a wide range
(e.g. \Td\ between 10 and 50 K).  For synthetic dataset A, we run the
chains for 18,000 iterations, and the traceplots show that after
$\sim$1000 draws the chains pass the $\hat{R}\approx 1$ convergence
test.\footnote{For the synthetic datasets A and B, each chain required
  approximately 1 hour on a single (2.5 GHz Intel) processor.}

The last two columns of Table \ref{bayes_DS_A} shows the posterior
means and 95\% highest posterior density (HPD) intervals.  Clearly,
the posterior means are very near the true underlying values, and the
HPD bracket the adopted values.  We have performed similar tests on
additional synthetic datasets, each with slightly different underlying
parameters.  Since the HPD of the posterior brackets the true value,
we can be confident of the HB model implementation in {\tt Stan}.

\begin{figure*}
\includegraphics*[width=140mm]{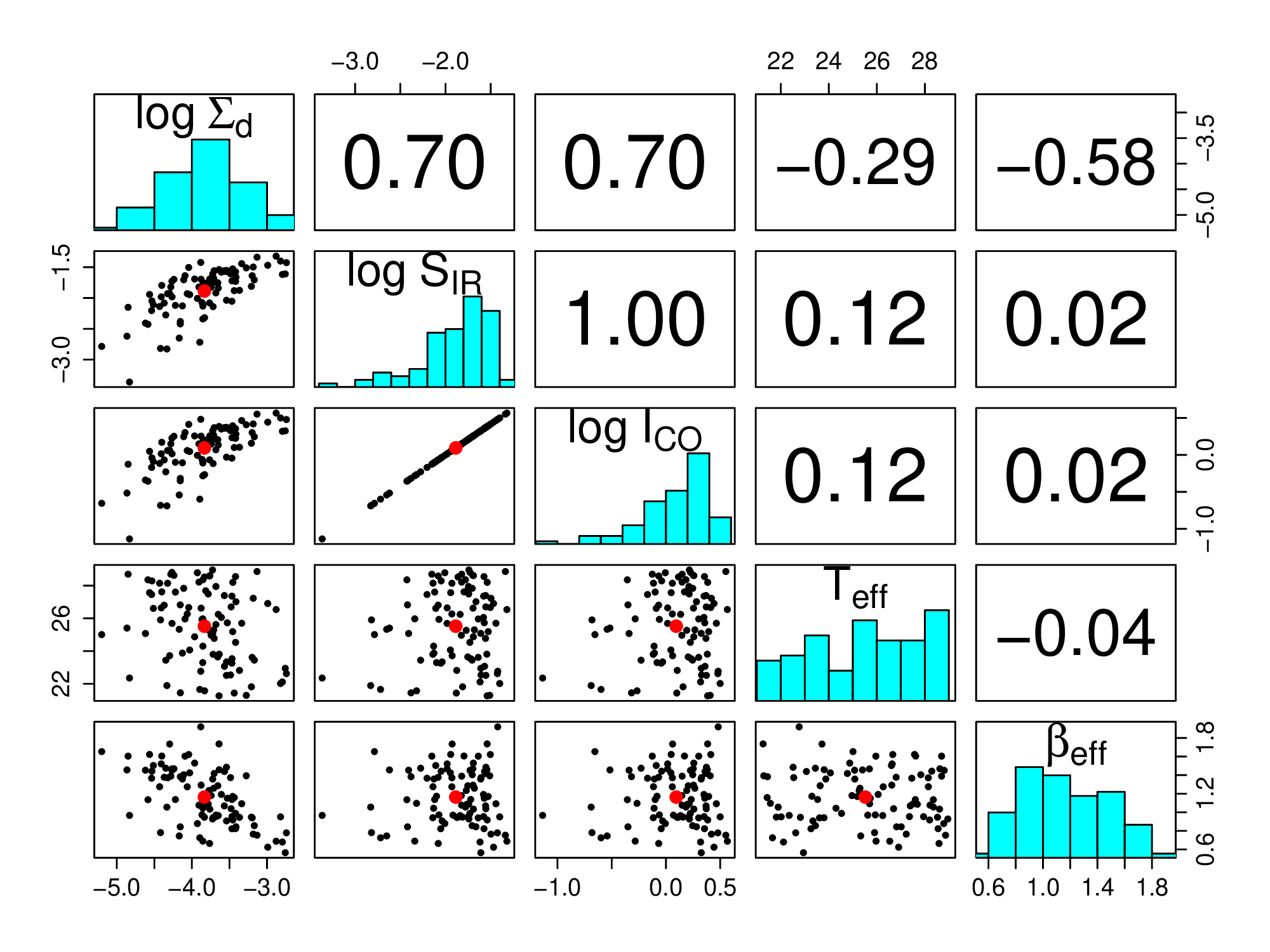}
\caption{Relationships between quantities in synthetic dataset B.  The
  panels are arranged in the same manner as Figure \ref{testA}.}
\label{testB}
\end{figure*}

In order to explore the effect of the prior distributions, we consider
datasets where the adopted distributions do not correspond to those
assumed in the HB model.  For this second test, we set \Ico\ to have a
truncated normal distribution.  As a result, $\log$\Ico\ is not
normally distributed, and thereby differs from the model assumption.
We choose a low value for $\overline{I}_{\rm CO}$ and a large
variance.  As we draw values for $I_{\rm CO, i}$, we discard any data
with $I_{\rm CO, i} < 0$.  This truncation thereby directly affects
the \Sir\ distribution.  Eliminating replicates with
$I_{\rm CO, i} < 0$ also modifies the other latent variables.  In
addition, we draw \Td\ from a uniform distribution, while keeping the
normal distribution of \betaeff.  With model misspecification, the
distributions of the latent variables deviate from their chosen prior
distributions in the HB model.  Figure \ref{testB} shows the
distributions and bivariate relationships of the underlying quantities
in synthetic dataset B, which consists of 100 datapoints.  We include
correlated calibration and random uncertainties: 10\% and 5\% for the
PACS bands, respectively, and 8\% and 7\% for the SPIRE bands.  It is
clear from Figure \ref{testB} that the synthetic quantities do not
follow the distributions adopted in the priors of the HB model.

We choose initial parameter values and chain lengths similar to those
employed for synthetic dataset A. We find that 18,000 iterations for
each of the three chains display convergence for all parameters,
including $\hat{R}\approx$1 and large effect sample sizes \apgt 4000.
Columns 1 - 4 of Table \ref{bayes_DS_B} lists the adopted values and
the HB estimated values and their 95\% HPDs of the key parameters in
synthetic dataset B.  As with test A, the 95\% HPDs of each parameter
includes the true values of dataset B, and the posterior means are
similar to the true values.

\begin{table*}
 \centering
 \begin{minipage}{140mm}
   \caption{Intrinsic, HB, and RIME (\chisq )
     Estimated Population Parameters for Synthetic Dataset B}
  \begin{tabular}{ccccc}
    \hline
    \hline
    Parameter & True Value & Posterior Mean & 95\% HPD & RIME Estimates \\
    \hline
    $\rho_{T,\beta}$ & 0.0 & -0.03 & [-0.32, 0.26] & $-0.22\pm0.1$ \\
    $\mu_{T_{\rm eff}}$ & 25.3 & 25.5 & [24.9, 26.1] & 25.7$\pm0.2$ \\
    $\sigma_T$ & 2.2 & 2.3 & [1.8, 2.8] & 2.2$\pm0.3$ \\
    $\mu_{\beta_{\rm eff}}$ & 1.24 & 1.18 & [1.10, 1.25] & 1.21$\pm0.03$ \\
    $\sigma_\beta$ & 0.33 & 0.31 & [0.25, 0.37] & 0.32$\pm0.03$ \\
    $\log \overline{I}_{\rm CO}$ & 0.15 & 0.09 & [0.02, 0.16] & - \\
    $\sigma_{\rm CO}$ & 0.28 & 0.32 & [0.28, 0.37] & - \\
    $n$ & 1.20 & 1.20 & [1.16, 1.25] & 1.14$\pm0.05$ \\
    $A$ & $-$2.00 & -1.99 & [-2.04, -1.94] & -1.99$\pm0.07$ \\
    \hline
\end{tabular}
\footnotetext[0]{$^1$ This dataset contains 100 repeated measures of
  CO and IR luminosities.  As for synthetic dataset A, the effective
  sample size $\approx$4000 and $\hat{R}$=1.}
\label{bayes_DS_B}
\end{minipage}
\end{table*}

We compare the HB results with parameters estimated via commonly
employed methods.  For the latter, we perform a simple regression
ignoring measurement error, or RIME, analysis of synthetic dataset B.
For each pixel, we fit the modified blackbody to the five intensities
by minimizing equation \eqref{chisqr-IR-px} and then perform linear
regression between the estimated $\log$\Sir\ and
$\log \hat{I}_{\rm CO}$, described by equation \eqref{chisqr-LR}.  We
simulate a large number of datasets with the same properties as
synthetic dataset B, including a correlated term for the calibration
uncertainty and a random noise term.  We then perform a RIME analysis
of each realization.  We compare the resulting sampling distributions
of the fit parameters to the estimates from the HB model.\footnote{We
  note that the sampling distributions of the estimated quantities
  should not correspond to the 95\% HPD of the HB results, as the
  range in posterior means would be a more suitable comparison.
  Nevertheless, we perform this comparison in order to obtain an
  estimate of the RIME uncertainty, and because similar analyses are
  commonly employed for ascertaining the uncertainties when fitting
  noisy observations.}

\begin{figure}
\includegraphics*[width=90mm]{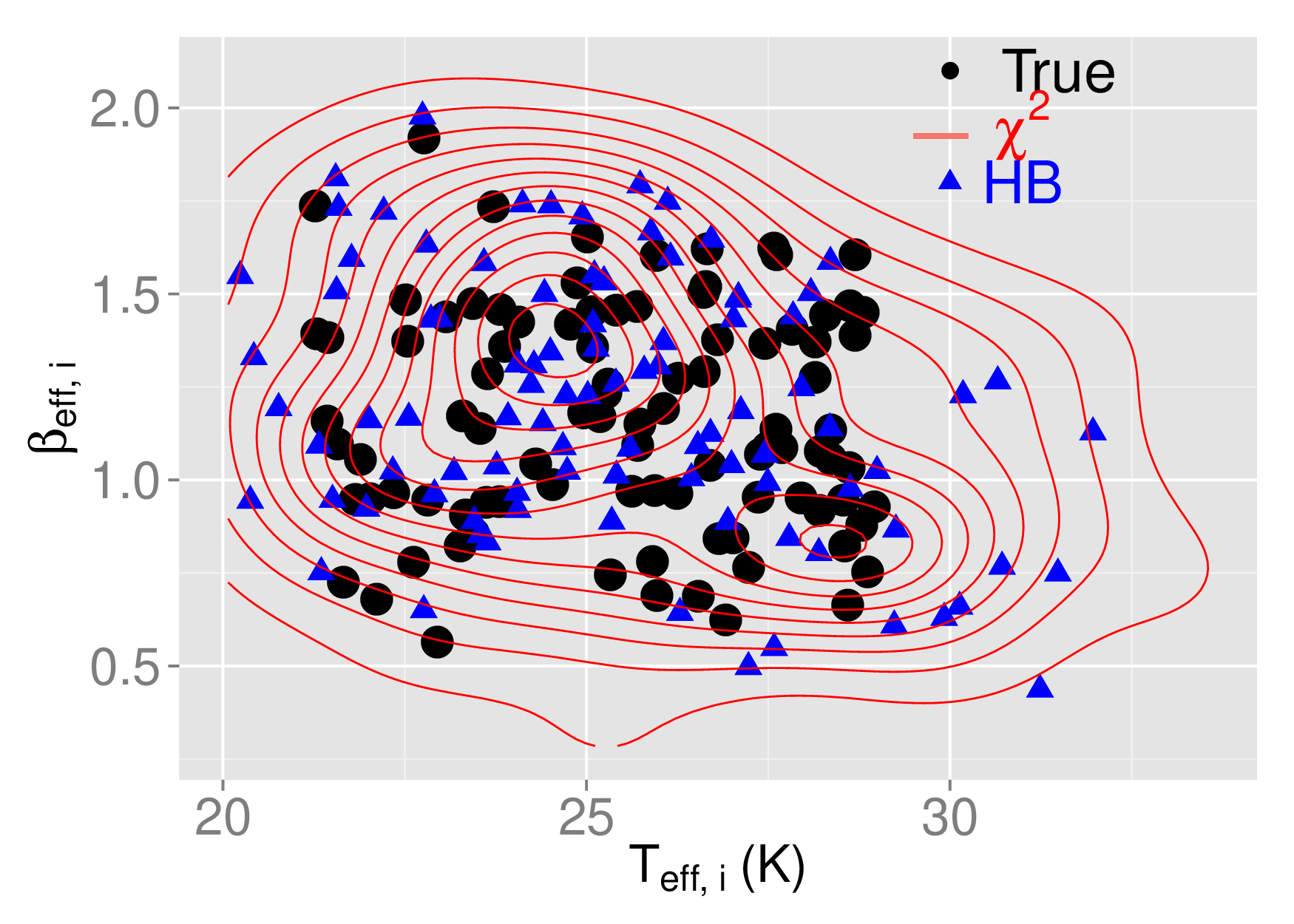}
\caption{Comparison of estimated \Td\ and \betaeff\ from the HB (blue)
  and \chisq\ point estimate methods (red contours, computed via
  kernel density estimation).  The black points show the true values
  in synthetic dataset B.}
\label{T_beta_fit}
\end{figure}

Figure \ref{T_beta_fit} shows the RIME estimates, HB posterior mean,
and the true values of \Td\ and \betaeff.  The last column of Table
\ref{bayes_DS_B} shows the best RIME parameter estimates and the
\tsig\ uncertainty of the underlying parameters from the sampling
distribution.  The RIME results indicate an anti-correlation between
\Td\ and \betaeff, with \rhoTB=$-$0.23.  Since the degeneracy between
\Td\ and \betaeff\ is explicitly treated in the HB method via the
modeling of \Td\ and \betaeff\ with a bivariate normal distribution,
the HB method is more reliable in estimating their correlation
\citep{Kelly+12}.

We can quantify the fits by computing the mean squared error (MSE) of
the RIME estimates and HB samples.  The MSE of \betaeff\ of both
methods are similar, 0.03.  However, the MSE of \Td\ from the RIME
method is 2.86, which is about 6\% higher than the mean MSE of 2.69
provided by the HB analysis.  Even though the HB estimate misspecifies
the population parameters, it is able to provide more accurate
estimates of \Td\ than the RIME point estimates.

\begin{figure}
\includegraphics*[width=90mm]{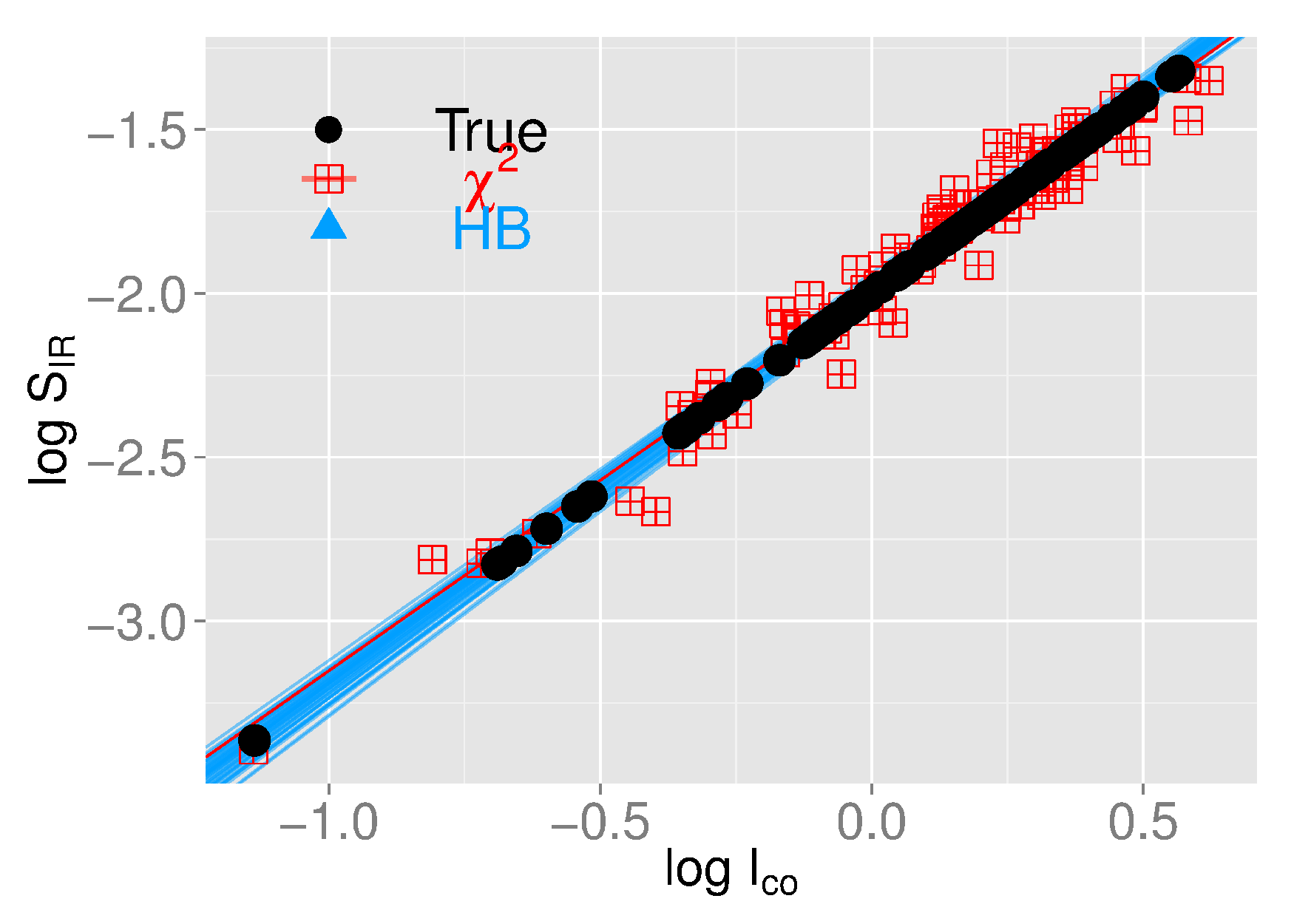}
\caption{Comparison of estimated and true \Ico\ $-$ \Sir\ relationships
  for test dataset B.  The blue lines are 20 random draws from the HB
  posterior, and the red points show the \chisq\ estimates of \Sir\ at
  the observed values of $\hat{I}_{\rm co}$, with the red line
  depicting the best fit.  The black points show the true values in
  synthetic dataset B.}
\label{Lir_Lco_fit}
\end{figure}

From the RIME estimates of \Td, \betaeff\, and \Sigd, we can integrate
the SED defined by these quantities to obtain estimates of \Sir.
Figure \ref{Lir_Lco_fit} shows the comparison of the true and fit
\Ico\ $-$ \Sir\ relationships.  The blue lines show 20 random draws
from the HB posterior.  The red points show the RIME estimated \Sir\
plotted along with the observed (i.e. noisy) $\hat{I}_{\rm CO}$
values.  This differs from the HB model, which explicitly estimates
the true values of \Ico; these values are used in the evaluation of
the \Ico\ $-$ \Sir\ relationship.  The RIME estimate of the slope of
\Ico\ - \Sir\ is 1.13, with \tsig\ uncertainty 0.05, as indicated in
Table \ref{bayes_DS_B}.  This value is flatter than the underlying
value $n$=1.2, even at the \tsig\ level.  As shown by numerous
previous works, when measurement uncertainties in the predictor are
not treated in the linear regression, the fit slope will be biased
towards 0 \citep[see e.g.][]{Akritas&Bershady96, Carrol+06, Kelly07}.
Note also that a linear regression on observations with relative IR
noise levels \apgt10\% will estimate a slope with more bias than the
results from synthetic dataset 2.  The HB estimate of $n$, on the
other hand, explicitly treats measurement uncertainties in all
observables, thereby leading to significantly more accurate parameter
estimate than the \chisq\ value.

We have also checked the relationship between each IR band and
$\hat{I}_{\rm CO}$.  Fits involving any individual IR intensities
usually underestimates the underlying slope, with the magnitude of the
bias depending on the noise characteristics, as well as the true
values of the parameters.  In general for higher noise levels, we have
found that the estimated slopes using solely the longer wavelength
intensities perform worse than those at shorter wavelengths.  Such
tests suggest that, similar to the RIME SED fits, employing any
individual intensity will likely lead to an underestimate of the slope
of the FIR $-$ CO relationship.

We have also performed a test on a synthetic dataset where the \Td\
$-$ \betaeff\ distribution is not bivariate normal.  As we show in the
next Section, for the LMC the \Td\ $-$ \betaeff\ distribution has
curvature, indicating that the model is mis-specified.  To assess
whether such mis-specification unduly influences the estimates of the
main parameters of interest, such as $n$, we construct a sythetic
dataset using the posterior mean of \Td\ and \betaeff\ from the LMC.
We create all the other latent parameters as in Synthetic Dataset B,
and perform the HB model of this synthetic dataset.  We find that the
95\% HPD brackets the true value of the underlying latent parameters.
A rigorous test of such mis-specification (e.g., verifying approximate
coverage of 95\% regions) would require the execution of the HB model
on thousands of such sythetic datasets.  Since this is unfeasible, our
limited tests considering only 5 synthetic datasets suggests that,
broadly speaking, the results are robust to mis-specifying the \Td\
and \betaeff\ distribution.

Before concluding this section, we stress that though the HB model
employs the simple assumption of a single (effective) temperature and
spectral index, the total FIR luminosity estimate is not significantly
affected by this constraint.  To verify this, we have constructed
simple models with an additional colder dust component, with
temperatures in the range 5 $-$ 15 K, and with surface densities that
are 5 $-$ 20 larger than the warm component, and fit the
single-component SED to the emergent IR intensities at the {\it
  Herschel} wavelengths.  Though the fit parameters can be discrepant
from true temperatures and spectral indices regardless of which
component dominates the emergent SED, the estimated integrated SED
recovers \Sir\ to a very high degree of accuracy, often within 0.5\%.
This demonstrates the robustness of \Sir, and that the fit temperature
and spectral index should only be interpreted as ``effective''
parameters.  Therefore, \betaeff, which determines the Rayleigh-Jeans
tail of the fit SED, must be understood only as the appropriate
numerical parameter required to reproduce the observed intensities
within the single-component SED model.  For an investigation focusing
on the properties of dust along each LoS, additional IR fluxes in the
Rayleigh-Jeans tail of the SED would be necessary to constrain the
\betaeff, and a multi-component fit may better recover the observed
trends.  In this study, we are primarily interested in the total IR
luminosity that is not strongly affected by the precise parameter
estimates of \betaeff, so we can be confident that the information
provided by the {\it Herschel} images is sufficient for estimating
\Sir.

In summary, the tests on synthetic data demonstrate that the HB model
performs well under certain conditions.  For synthetic dataset A the
underlying distributions matched those assumed in the HB model,
resulting in a highly accurate and precise posterior.  Synthetic
dataset B includes variables with distributions which are discrepant
from those assumed in the HB model.  Nevertheless, the HB method is
able to recover the underlying latent parameters within the 95\% HPD.
We note that a RIME fit to the intensities of synthetic dataset B
recovers the mean values and marginal distributions of \Td, \betaeff\,
and \Sir.  However, both \rhoTB\ and $n$ are significantly
underestimated.  Such discrepancies advocate for employing a
hierarchical modelling technique, which relaxes the strong assumption
of independence between \Td\ and \betaeff\ adopted in the RIME fit.

\section{Application: The Large and Small Magellanic Clouds} \label{datasec}
\subsection{Observations}

\begin{figure*}
\includegraphics*[width=140mm]{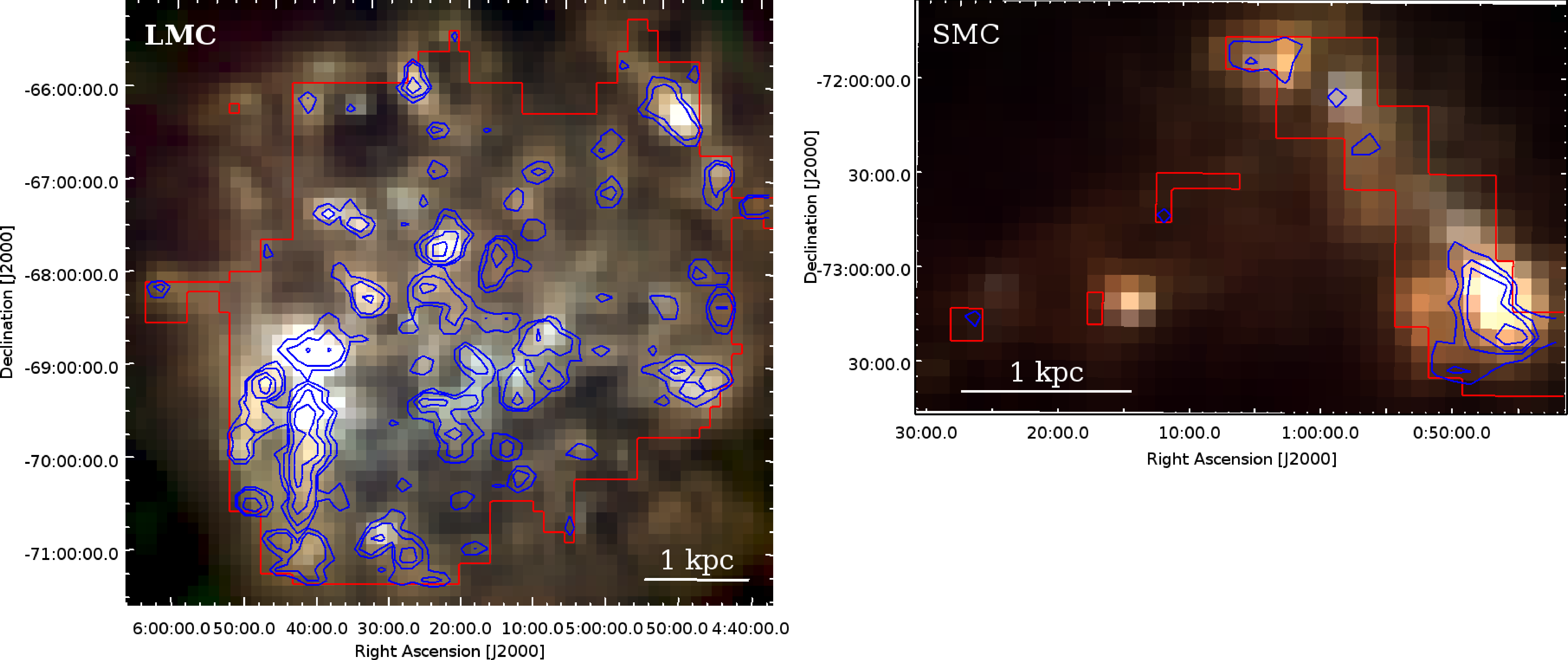}
\caption{Three-color image of the LMC and SMC at 100 pc scale from the
  100, 160, and 250 \micron\ {\it Herschel} observations.  The blue
  contours show the CO intensities at by 0.2, 0.3, 0.5, 1 and 2
  \Icounits\ for the LMC, and 0.1, 0.2, 0.3, 0.5 and 1 \Icounits\ for
  the SMC. The red box demarcates the observed region of the CO maps}
\label{LMCimg}
\end{figure*}

As an initial application of the HB method on real data, we analyze
the dust and gas properties of the Magellanic Clouds using the {\it
  Herschel} HERITAGE key project FIR data \citep{Meixner+13}, in
conjunction with CO $J=1-0$ NANTEN observations \citep{Mizuno+01a,
  Mizuno+01b}.  We convolve all maps to a common resolution of 100 pc,
since we are interested in the large scale properties of the ISM.  We
consider those pixels with \Ico\ $> 0$ \Icounits.  Given these
constraints, the number of independent pixels we analyze in the SMC
and LMC is 132 and 1584, respectively.  To illustrate the maps we
employ for this work, Figure \ref{LMCimg} shows the LMC and SMC map at
100 pc scales from the 100, 160, and 250 \micron\ observations, with
the contours displaying the CO peaks.

We have quantified the spatial correlation of residuals between
neighbouring pixels as a result of the beam convolution at the 100 pc
resolution we employ.  We have measured the correlation coefficient of
the residual values of neighbouring 100$\times$100\,pc$^2$ pixels in
synthetic data with similar resolution and gridding to {\it Herschel}
images.  We find that the residuals can be correlated by upto $\sim$20
per cent between adjacent pixels.  We have checked that this
correlation likely does not influence our results as we have run the
HB model on every other pixel of the LMC maps, in which case the
employed pixels are uncorrelated, and have recovered similar results
to that from the full maps.  Therefore, the $\sim$20 per cent
introduced correlation between adajacent pixels is sufficiently small
such that we can confidently apply the HB model, assuming conditional
independence between pixels.

\subsection{Uncertainties and Photometric colour corrections} \label{uncsec}

The HB model requires (fixed) values for the standard deviations of
the correlated and uncorrelated uncertainties discussed in Section
\ref{mmsec} .  Following \citet{Muller+11}, we set the following for
the PACS (j=1 or 2, corresponding to 100 and 160 \micron)
uncertainties:
\begin{eqnarray}
\mathcal{C_{\rm PACS}} \sim \mathcal{N}(0, \log(1.1)) \label{Cpacs}  \\
\sigma_{\rm j={1\, or\, 2}} = \log(1.02). \label{sigpacs}
\end{eqnarray}
As reported in \citet{Griffin+13} and \citet{Bendo+13}, for SPIRE
(j=3, 4, or 5, corresponding to 250, 350, and 500 \micron) we set
\begin{eqnarray}
\mathcal{C_{\rm SPIRE}} \sim \mathcal{N}(0, \log(1.08))    \label{Cspire}  \\
\sigma_{\rm j={3,\, 4\, or 5}} = \log(1.015). \label{sigspire}
\end{eqnarray}

To compare the model SEDs with the observations, the analytic modified
blackbody (MBB) curves should be convolved with the PACS and SPIRE
filter response
functions\footnote{\url{http://herschel.esac.esa.int/Docs/PACS/html/pacs_om.html},
  \url{http://herschel.esac.esa.int/Docs/SPIRE/html/spire_om.html}}
to produce synthetic {\it Herschel} photometry.  In order to speed up
the calculations and because the corrections only depend on the shape
of the SED, we pre-calculate colour corrections ($CC$) factors for
MBBs with a range of $T$ and $\beta_{\rm eff}$ values. The
correction factors are defined as F$_{{\nu},\mathrm{Herschel}}$ =
$CC_{\nu} \cdot$F$_{{\nu},\mathrm{mono}}$, where
F$_{{\nu},\mathrm{Herschel}}$ is the flux-density value as measured by
the {\it Herschel} bolometers and F$_{{\nu},\mathrm{mono}}$
is the monochromatic flux-density, i.e. the value in the perfectly
sampled model SED, at the reference wavelength.\footnote{We follow
  \url{http://herschel.esac.esa.int/twiki/pub/Public/PacsCalibrationWeb/cc_v1.pdf}.}
We calculate $CC$ values for 15\,K$\leq T\leq$40\,K and
0$\leq \beta_{\rm eff}\leq$3.  In general the correction factors are
relatively small but significant. Over the immediate range of interest
(20\,K$<$T$<$30 K and 0.5$<\beta_{\rm eff}<$2) the maximum amplitude
($\|CC-1\|$) of the correction is 0.02, 0.05, 0.05, 0.07 and 0.11
for the 100, 160, 250, 350 and 500 $\mu$m filters, respectively.

Inside {\sc stan}, the colour correction factors are specified as a
second order polynomial in $T$ and \betaeff\ whose parameters have
been determined by fitting a 2D plane to the $CC$ values using the
{\sc IDL} routine SFIT.

\subsection{Results} \label{ressec}

\begin{figure}
\includegraphics*[width=90mm]{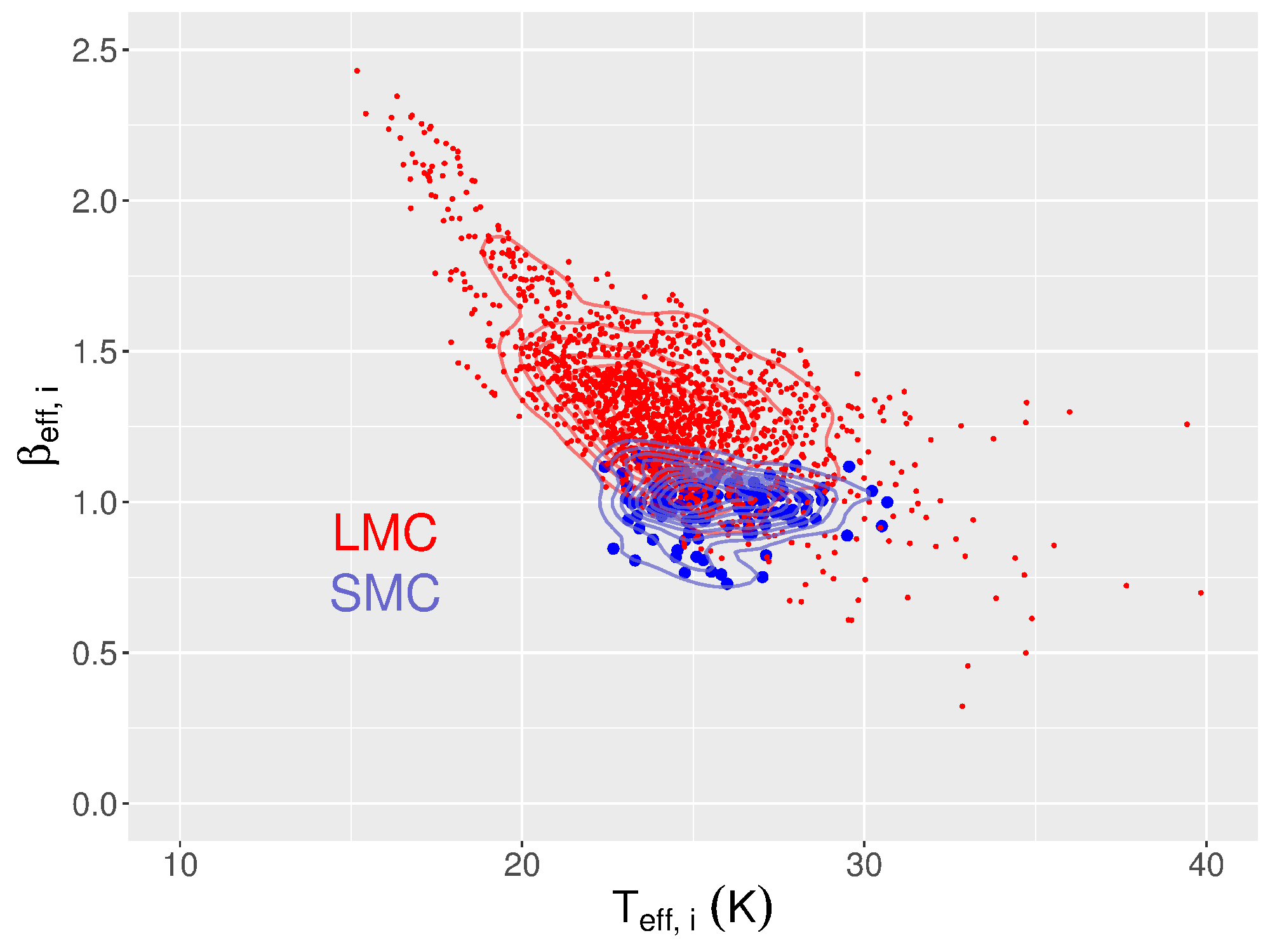}
\caption{The \Td $-$ \betaeff\ relationships for the LMC (red) and SMC
  (blue).}
\label{both_T_beta}
\end{figure}

Figure \ref{both_T_beta} shows the marginal means of the estimated
\Td\ and \betaeff\ in each pixel of the SMC and LMC.  Tables
\ref{LMC_TB_TAB} and \ref{SMC_TB_TAB} display the posterior mean and
95\% HPDs of the location and scale parameters of the \Td\ $-$
\betaeff\ distributions.  We find that the distributions of \Td\ and
\betaeff\ are rather different between the two galaxies, with the LMC
showing a larger anti-correlation (mean value
$\overline{\rho}_{T,\beta}=-0.74$) than the SMC
($\overline{\rho}_{T,\beta}=-0.15$).  In general, the galaxies also
have different mean temperatures and spectral indices location
parameters, with $\mu_{T_{\rm eff}}\approx$ 24.3 in the LMC and
$\approx$ 25.7 K in the SMC.  The $\mu_{\beta_{\rm eff}} \approx$1.32
for the LMC, and $\approx$0.99 for the SMC.

\begin{table}
 \centering
 \begin{minipage}{140mm}
  \caption{HB Estimated \Td\ - \betaeff\ parameters for the LMC}
  \begin{tabular}{ccc}
  \hline
  \hline
 Parameter & Posterior Mean & 95\% HPD \\
\hline
$\rho_{T,\beta}$ & -0.74 & [-0.76, -0.7]  \\
$\mu_{T_{\rm eff}}$ & 24.3 & [24.1, 24.5]  \\
$\sigma_T$ & 4.3 & [4.1, 4.4]  \\
$\mu_{\beta_{\rm eff}}$ & 1.32 & [1.31, 1.33]  \\
$\sigma_\beta$ & 0.28 & [0.27, 0.29]  \\
\hline
\end{tabular}
\label{LMC_TB_TAB}
\end{minipage}
\end{table}

\begin{table}
 \centering
 \begin{minipage}{140mm}
  \caption{HB Estimated \Td\ - \betaeff\ parameters for the SMC}
  \begin{tabular}{ccc}
  \hline
  \hline
 Parameter & Posterior Mean & 95\% HPD \\
\hline
$\rho_{T,\beta}$ & -0.15 & [-0.34, 0.03]  \\
$\mu_{T_{\rm eff}}$ & 25.7 & [25.4, 26.0]  \\
$\sigma_T$ & 1.8 & [1.6, 2.0]  \\
$\mu_{\beta_{\rm eff}}$ & 0.99 & [0.97, 1.00]  \\
$\sigma_\beta$ & 0.1 & [0.09, 0.12]  \\
\hline
\end{tabular}
\label{SMC_TB_TAB}
\end{minipage}
\end{table}

As is evident from Figure \ref{both_T_beta}, the \Td\ $-$ \betaeff\
distribution is not bivariate normal, especially for the LMC.  This
indicates that this aspect of the HB model is mis-specified, and the
HB parameter \Td\ $-$ \betaeff\ estimates do not have a
straightforward interpretation as moments of the population
distribution; rather, they are moments of a bivariate normal
approximation to that distribution.  Nevertheless, from the simulation
studies described in Section 3, we can be confident that the other
estimated parameters are not strongly affected by this specification.
One possible reason for the curved \Td\ $-$ \betaeff\ relationship is
that the single-temperature SED does not accurately reproduce the
shape of the observed spectrum, as we further discuss below.

\begin{table}
 \centering
 \begin{minipage}{140mm}
  \caption{HB Estimated Parameters for the LMC}
  \begin{tabular}{ccc}
  \hline
  \hline
 Parameter & Posterior Mean & 95\% HPD \\
\hline
$\log \overline{I}_{\rm CO}$ & -0.59 & [-0.61, -0.57]  \\
$\sigma_{\rm CO}$ & 0.32 & [0.28, 0.36]  \\
$n$ & 1.25 & [1.11, 1.38]  \\
$A$ & $-$2.5 & [$-$2.6, $-$2.4]  \\
\hline
\end{tabular}
\label{LMC_TAB}
\end{minipage}
\end{table}

\begin{table}
 \centering
 \begin{minipage}{140mm}
  \caption{HB Estimated Parameters for the SMC}
  \begin{tabular}{ccc}
  \hline
  \hline
 Parameter & Posterior Mean & 95\% HPD \\
\hline
$\log \overline{I}_{\rm CO}$ & -0.76 & [-0.82, -0.69]  \\
$\sigma_{\rm CO}$ & 0.21 & [0.17, 0.26]  \\
$n$ & 1.47 & [1.21, 1.72]  \\
$A$ & $-$2.1 & [$-$2.3, $-$2.0]  \\
\hline
\end{tabular}
\label{SMC_TAB}
\end{minipage}
\end{table}

The HB estimates of \Td\ and \betaeff\ in the Magellanic Clouds are
comparable to previous estimates.  \citet{Aguirre+03} also found that
the mean spectral index in the SMC is lower than in the LMC.  The
absolute values of their estimates differ from those we estimate here,
in part because they analyze all emission from the galaxies, whereas
we only consider dense regions where CO is detected.

\begin{figure}
  \includegraphics*[width=90mm]{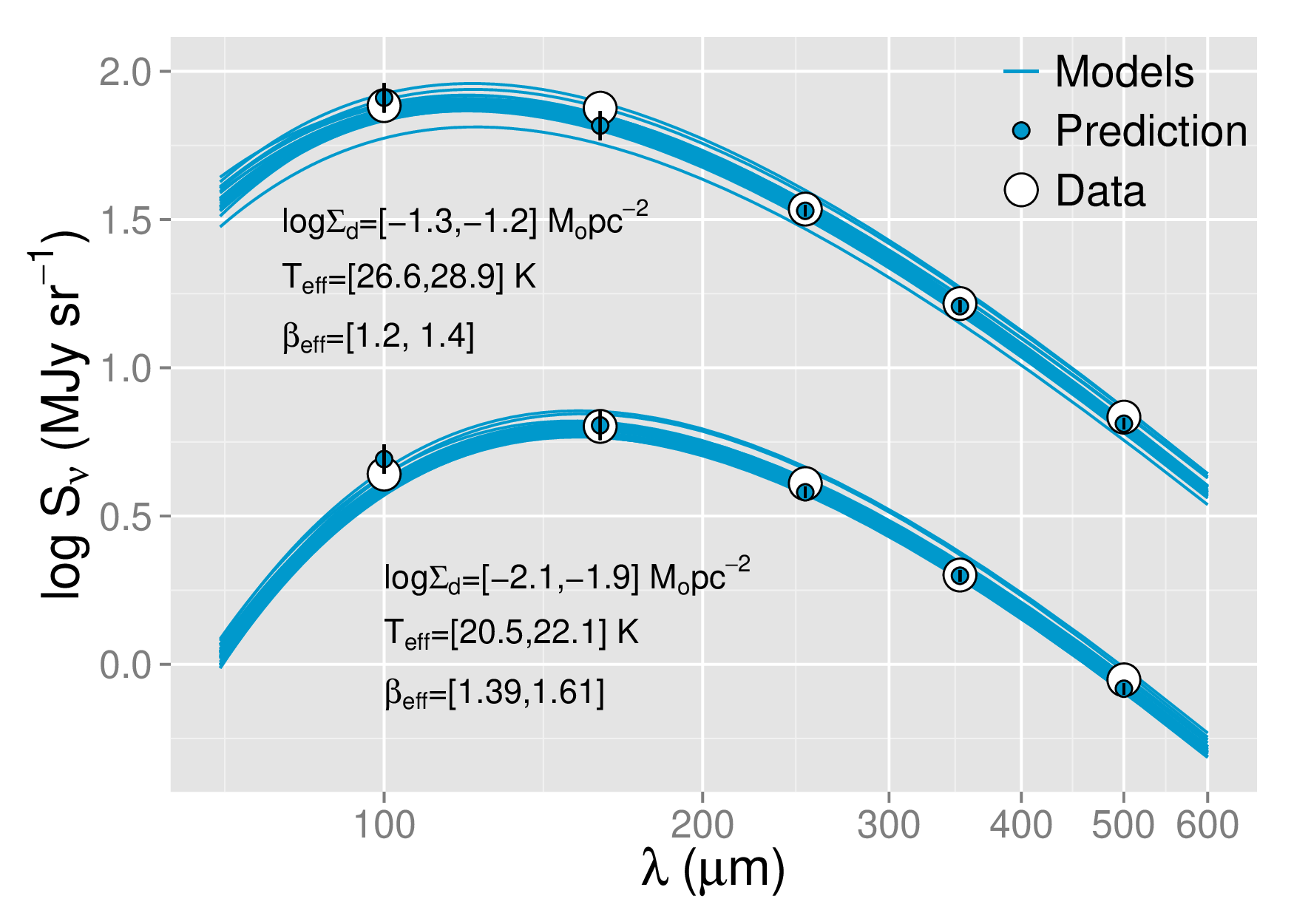}
  \caption{Example SED fits of two pixels in the LMC.  These pixels
    correspond to regions that have the largest and smallest \Sir.
    The lines are random draws from the posterior, and the blue points
    with errorbars show the mean {\it Herschel} predicted intensities
    with \tsig\ uncertainties.  The slight offsets between the
    predictions and the model SEDs are due to the calibration
    uncertainties and color corrections.  The white circles show the
    observed intensities ($\hat{S}_{\rm i, j}$).  The legend below
    each set of SEDs shows the 95\% HPDs of the fit parameters.}
\label{LMC_sed}
\end{figure}

\begin{figure}
  \includegraphics*[width=90mm]{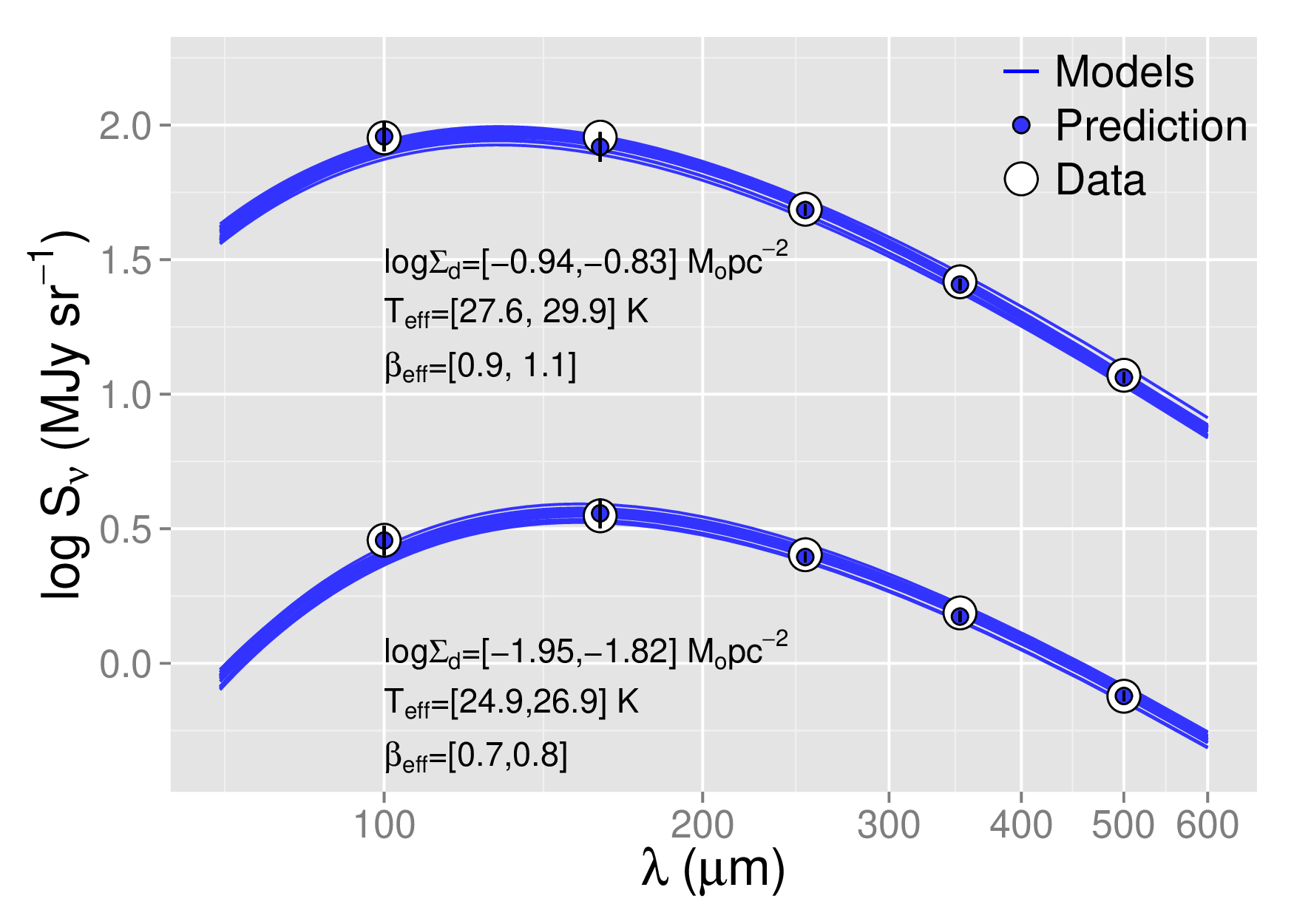}
  \caption{SED fits in pixels among the highest and lowest \Lir\ in
    the SMC.  The model, predictions, and observed data are marked in
    the same fashion as Figure \ref{LMC_sed}.}
\label{SMC_sed}
\end{figure}

To illustrate that the HB modelling has accurately reproduced the
observed data, we show two sample SEDs from each galaxy in Figures
\ref{LMC_sed} and \ref{SMC_sed}.  The models and data correspond to
the regions with some of the highest and lowest \Sir\ from each
galaxy.  The figures show the observed IR intensities, as well as the
predicted intensities at the corresponding wavelengths.  Note that the
offset between the prediction and data from the model SEDs, especially
at the highest frequencies, is expected, primarily due to the color
corrections described in Section \ref{uncsec}, and to a lesser extent
the correlated uncertainties.  The observations fall within the
expected range of the predicted intensities, indicating that the HB
method accurately recovers the observed data.

The differences in \rhoTB\ may be due to a number of possible
variations in the structure of the ISM between the SMC and LMC .  For
instance, temperature gradients along the LoS may induce spurious
correlations \citep[e.g.][]{Shetty+09b, Malinen+11, JuvelaYsard12,
  Veneziani+13, Gordon+14}.  There may be contrasting \Td\ gradients,
either in extent or magnitude, in the Magellanic Clouds leading to the
difference in \rhoTB.  Other possibilities are that the dust-to-gas
ratio and 500 \micron\ excess may vary\footnote{However, we have
  checked that excluding the 500 \micron\ observations from the fits
  in the SMC does not significantly alter the HB estimated
  parameters.}\citep[e.g.][]{Bernard+08, Roman-Duval+10a, Gordon+14}.
Though measurement uncertainties can produce an artificial
anti-correlation due to the degeneracy between \Td\ and \betaeff,
\citep[e.g.][]{Blain+03, Dupac+03, Shetty+09a}, the HB method
explicitly models the \Td$-$\betaeff\ correlation and accurately
accounts for noise, thereby reducing this effect
\citep[see][]{Kelly+12}.  Other violations of model assumptions may
also produce an artificial anti-correlation, such as a multiple dust
components along the LoS leading to SEDs with a broken power-law
Rayleigh-Jeans tail, and may be responsible for the apparent deviation
of the LMC points from a bivariate normal distribution in Figure
\ref{both_T_beta}.  As we have employed the same assumptions in
modelling both Magellanic Clouds, the estimated differences in the \Td
$-$ \betaeff\ indicates the presence of fundamental differences in the
dust characteristics of the LMC and SMC.

\begin{figure}
  \includegraphics*[width=90mm]{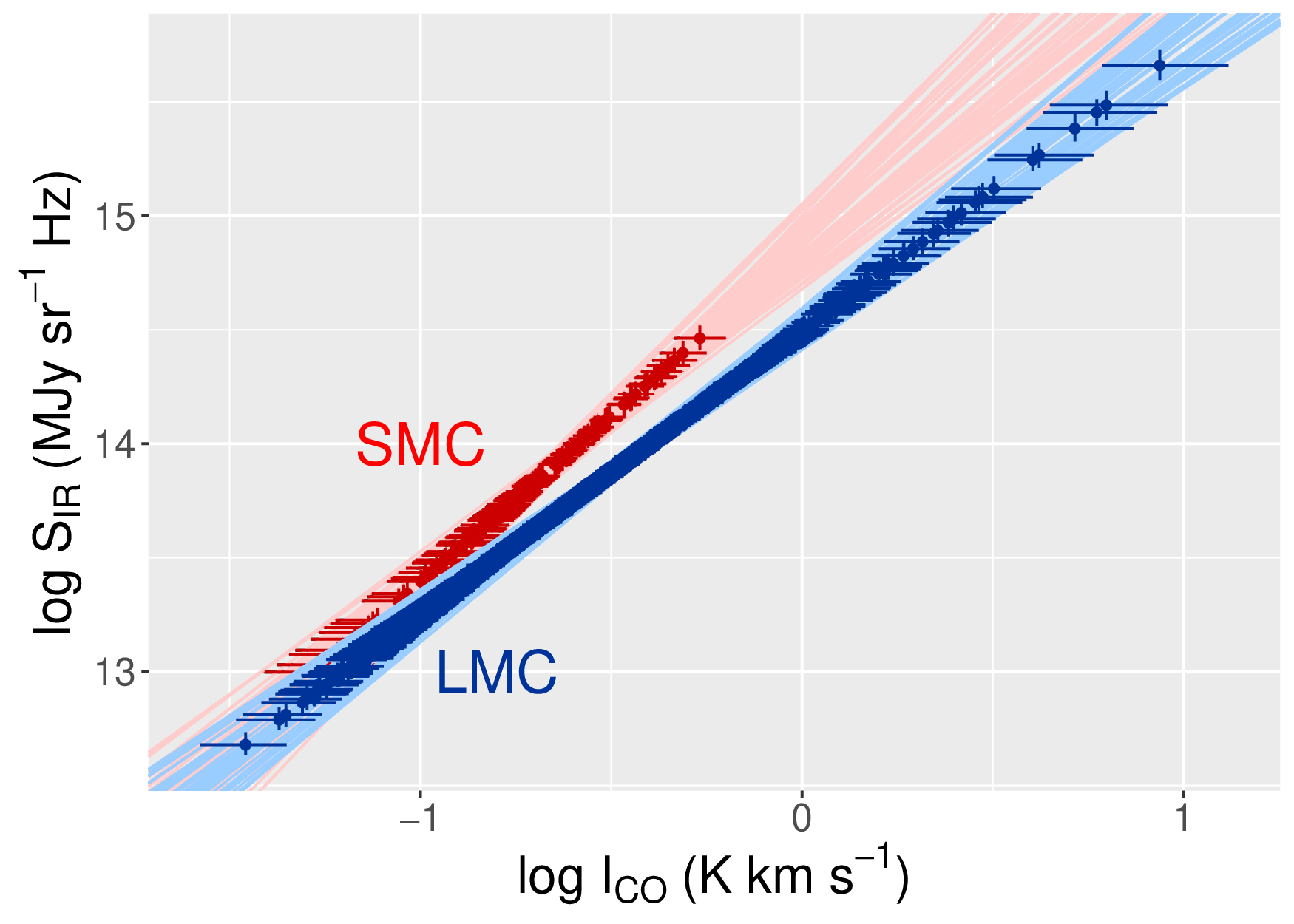}
  \caption{\Ico $-$ \Sir\ relationship in the LMC and SMC (50 random
    draws from the posterior).  Colored lines depict draws of
    regression lines from the marginal posterior distribution for
    $(A,n)$ for the SMC (red) and LMC (blue).  Points with error bars
    show the marginal posterior means and marginal (1-D) 95\% HPD
    regions for $(\log I_{\rm CO}, \log S_{\rm IR})$ for each pixel in
    the two images.}
\label{LMC_SMC}
\end{figure}

Tables \ref{LMC_TAB} and \ref{SMC_TAB} display the HB estimated
parameters related to the \Sir\ $-$ \Ico\ of the LMC and SMC.  Figure
\ref{LMC_SMC} shows 50 random draws of the \Ico$-$\Sir\ relationship
for both galaxies\footnote{As the regression lines all fall within the
  errors, we can be confident that no additional scatter is required
  to model the \Ico$-$\Sir\ relationship.}.  The slopes of the two
galaxies are similar; the posterior mean value of $n$ is 1.3 and 1.5
for the LMC and SMC, respectively, with overlap in the 95\% HPDs.  On
the other hand, the intercepts are clearly different.  Per unit \Ico,
the SMC has a $\sim$0.4 dex higher \Sir\ than the LMC.  This
difference is likely associated with the lower metallicity of the SMC,
0.2 \Zsun, compared to 0.5 \Zsun\ of the LMC.  If \Sir\ is a faithful
tracer of the star formation rate, these results suggest that the star
formation rate per unit \Ico\ is higher in the SMC by the same
$\sim$0.4 dex factor.  Numerous previous observations have shown such
trends for lower metallicity systems \citep[e.g.][]{Taylor+98,
  Bolatto+11, Leroy+07, Cormier+14}.  The salient difference is in the
ability to trace the star forming ISM with CO.  In lower metallicity
environments, there is a dearth of CO, so a given \Ico\ would be
associated with a higher SFR compared to the higher metallicity case
\citep{Maloney&Black88, Wolfire+10,
  Glover&MacLow11,Shetty+11a,Shetty+11b, Glover&Clark12,
  Clark&Glover15}.  Additionally, regions with higher atomic
densities, will contain more dust.  Therefore, dust emission should be
correlated with total (atomic + molecular) gas density, and the
offsets between the galaxies in Figure \ref{LMC_SMC} may be partially
due to the variations in total gas density, in addition to the
metallicity differences.

\begin{figure}
  \includegraphics*[width=90mm]{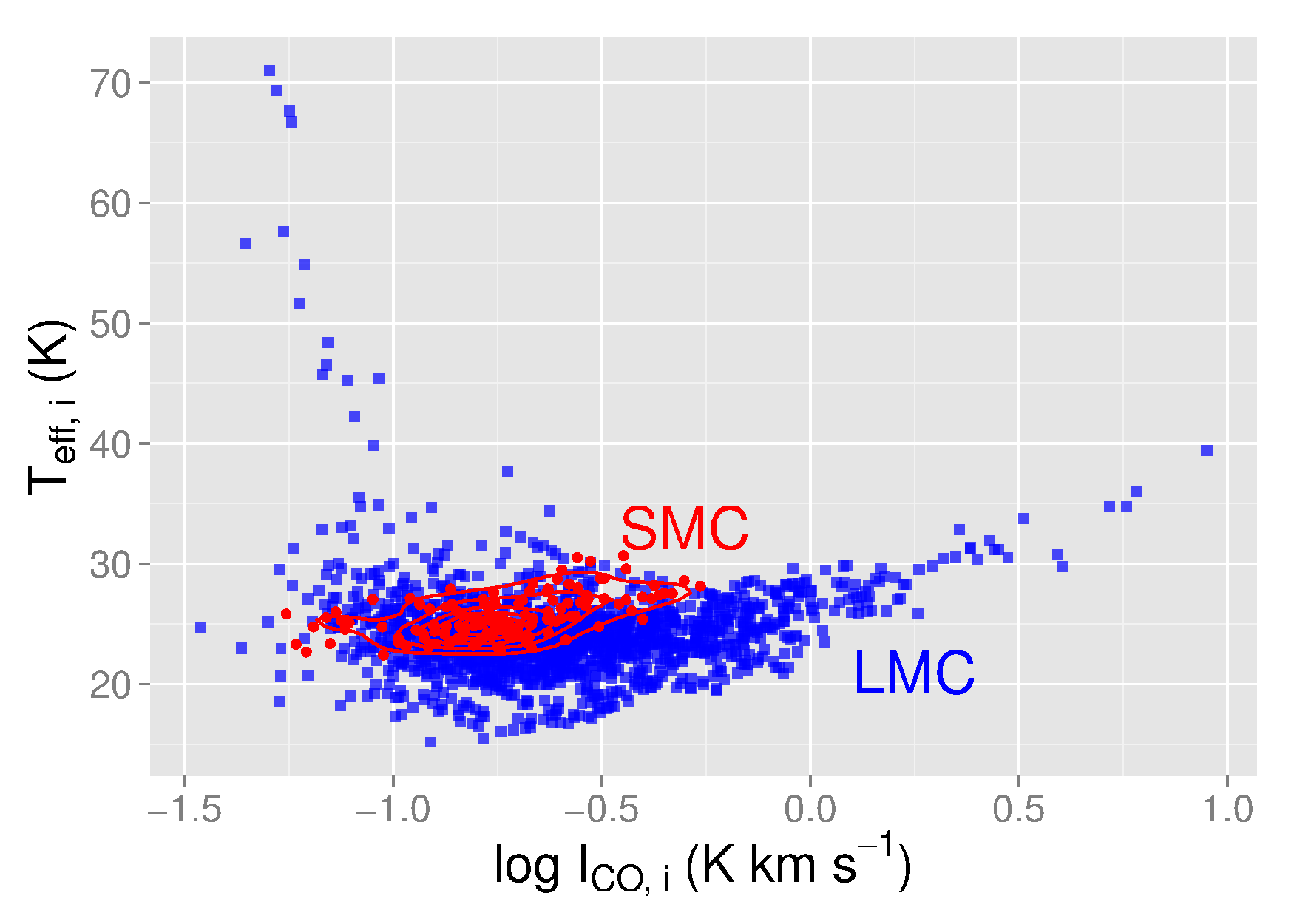}
  \caption{\Ico $-$ \Td\ relationship in the LMC (blue) and SMC (red).
    In order to better reveal the contrast between the Magellanic
    Clouds, we do not show contours for the LMC data.  The points show
    \Ico\ and \Sir\ at each pixel from the mean of the posterior.}
\label{LMC_SMC_TLco}
\end{figure}

The mean dust temperature is clearly larger in the SMC than the LMC,
contributing to the overall increase in \Sir\ (at a given \Ico).  Such
higher temperatures are observed in other lower metallicity systems
\citep[e.g.][]{Remy-Ruyer+13}.  Figure \ref{LMC_SMC_TLco} shows the
posterior mean relationship between \Td\ and \Ico.  In the LMC, at
large \Ico\apgt0.5 \Icounits, there is strong evidence that \Td\
increases from $\approx$ 20 K to $>$30 K where \Ico\ $\approx$ 3
\Icounits.  This increase in dust temperature is likely due to
radiation from young stars embedded in the dense molecular medium.  At
lower densities, we also find high \Td\ values.  Although this may be
due to a decline in self-shielding, in the regions with the lowest
\Ico\ the signal is close to the noise levels, so we should be careful
not to over-interpret these data.  At intermediate \Ico, there is a
large range in \Td, which is indicative of a mix of dense star forming
regions, diffuse gas, and all material at intermediate densities.  The
SMC does not portray any strong trends as the \Td\ estimates span a
smaller range, though there is some hint of an increase in \Td\ at the
highest \Ico.

\begin{figure*}
  \includegraphics*[width=90mm]{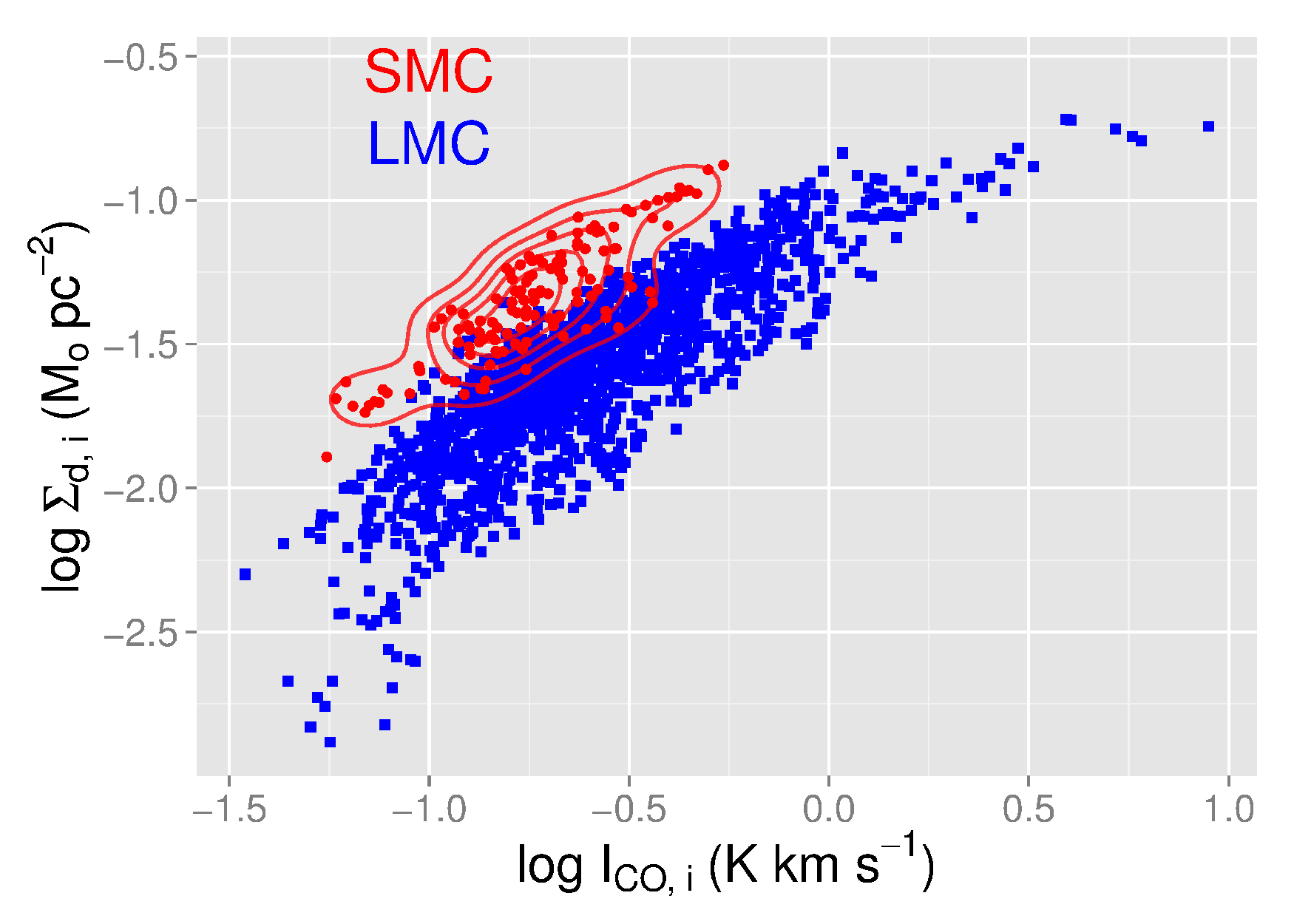}
\caption{Example \Ico $-$ \Sigd\ relationship in the LMC and SMC.}
\label{LMC_SMC_Nd}
\end{figure*}

One question is to what extent the dust surface density estimates
affect \Sir.  Figure \ref{LMC_SMC_Nd} shows the posterior mean of the
\Sigd\ $-$ \Ico\ trends in the SMC and LMC.  Again, there is an
unambiguous offset, with the SMC showing larger \Sigd\ per unit \Ico.
As discussed, the lower metallicity in the SMC results in fewer CO
molecules for a given dust surface density.  Therefore, it is the
combination of both higher temperatures and higher surface densities
per unit CO intensity, both due to the lower metallicity in the SMC,
which leads to the higher normalisation in the \Sir $-$ \Ico\ trend
shown in Figure \ref{LMC_SMC}.

\begin{figure*}
\includegraphics*[width=140mm]{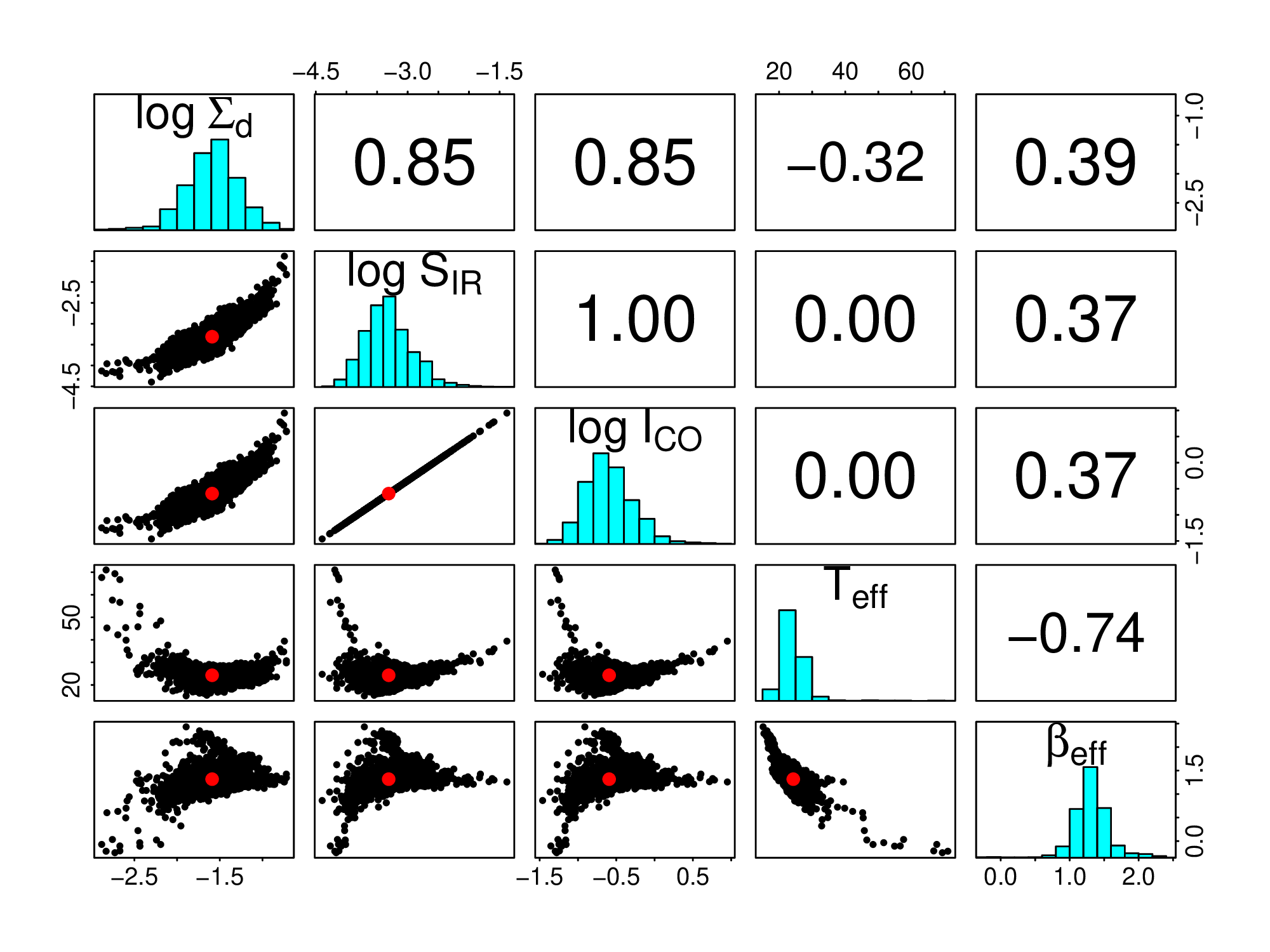}
\caption{Modelled relationships and distributions between the latent
  variables in the LMC.  The panels are arranged in the same manner as
  Figure \ref{testA}.}
\label{lmcsc}
\end{figure*}

The HB model provides estimates of all the latent variables that
determine the observed CO and IR intensities.  Figure \ref{lmcsc}
shows the modeled distributions and bivariate correlations of these
variables from the LMC.  The displayed model corresponds to the
posterior mean.  Though there are clearly strong trends between many
of the variables besides the bivariate relationships discussed above,
as we discuss in the next section, interpreting such correlations
requires a complete consideration of the key assumptions that enter
the model.

\section{Discussion} \label{discsec}

The HB modeling of dust SEDs and the dust - molecular gas relationship
on 100 pc scales has revealed some interesting comparative features of
the LMC and SMC.  There is a strong anti-correlation between \Td\ and
\betaeff\ in the LMC, with the SMC portraying a weaker
anti-correlation, which may even be uncorrelated.  Further, the slopes
of \Ico $-$ \Sir\ relationship are nearly identical, while \Sir\ in
the SMC is larger by a factor of $\sim$3 at any given \Ico.  These
results suggest that there are some fundamental differences in the
properties of the dust and gas between the galaxies.

Though the estimated \rhoTB\ is significantly different between the
LMC and SMC, we should be careful to draw too strong a conclusion from
this result.  Since the SED model does not account for multiple
components, the estimated \betaeff\ does not necessarily correspond to
the underlying spectral indices of any of the components along the
LoS.  Rather, it should be considered a convenient numerical parameter
necessary for fitting the SED model to the observations.  As \Sir\ is
not sensitive to the estimate of \betaeff, the modelled \Ico $-$ \Sir\
also does not strongly depend on \betaeff.  The SED modelling results
indicate that there is some fundamental difference in the properties
of dust, but further analysis is necessary to identify the cause of
these differences.  Possible causes may be associated with differences
in the magnitude of the 500 \micron\ excess, which is not modelled
here, the dust-to-gas ratio \citep[e.g.][]{Gordon+14, Roman-Duval+14},
and/or the dominance of the trend of some subset of the data (e.g. at
low or high \Td).  We have performed the HB fit to the SMC data
excluding the 500 \micron\ observation, but the resulting correlation
and distributions are similar, suggesting that any 500 \micron\ excess
is not unduly affecting the estimated correlations in the high density
regions traced by CO.  More in-depth analysis of the observed IR maps
may favor one of these or perhaps even other causes for the contrast
in \rhoTB, and whether the metallicity difference between the LMC and
SMC is a significant contributing factor.

That the estimated value of $n$ in both galaxies are (approximately)
equivalent may be indicative of a key similarity in the LMC and SMC.
If \Sir\ is considered to be a linear tracer of \sigsfr\, and \Ico\ is
a faithful tracer of \sigmol, then the HB results suggests that the
increase in the rate of star formation towards denser regions is
similar in the LMC and SMC despite their difference in metallicity.

Analyses of the KS relationship on large scales in normal disk
galaxies have revealed a range in $n$, with many galaxies portraying a
sub-linear \sigsfr $-$\sigmol\ relationship \citep[e.g.][]{Blanc+09,
  Ford+13}.  In fact, when using the single 24 \micron\ intensity or
its combination with UV observations, many galaxies in the HERACLES
and STING surveys favor a sub-linear KS relationship, with significant
galaxy-to-galaxy variations \citep{Shetty+13, Shetty+14a}.
\citet{Shetty+14b} interpret these and other observational results as
evidence for a substantial amount of non star-forming molecular gas
traced by CO, with this diffuse gas fraction consisting of at least
30\% of the total molecular content \citep[see also][]{Wilson+12,
  Caldu-Primo+13, Pety+13}.  Recent analyses of the Milky Way further
supports the presence of a significant diffuse molecular component
\citep[]{Roman-Duval+15, Liszt+10}.  Most of the
extra-galactic results utilized monochromatic SFR tracers, which could
contribute to the differences between the estimated indices from those
investigations and those presented here.

Another explicit difference is that the metallicities of the
Magellanic Clouds are significantly lower than those of the normal
spiral galaxies.  That the Magellanic Clouds have a super-linear KS
relationship may be due to their lower metallicity.  In low
metallicity systems, higher levels of photodissociation results in a
dearth of CO traced diffuse molecular gas compared to the normal
spirals \citep[e.g.][]{Cormier+10, Schruba+12, Glover&Clark12}.  Most
regions with sufficient amounts of CO coincide with the locations of
star formation.  Applying the HB model to extra-galactic surveys will
further reveal any trends between the KS index and other galaxy
properties, and will allow for a more direct comparison with the
results presented here.  Worthwhile comparisons of systems with
diverse metallicities involving dust SEDs would also require
consideration of the fraction of young stellar radiation that does not
heat the dust, which is estimated to be higher in the Magellanic
Clouds \citep[see][]{Lawton+10}.

Such investigations on a larger sample of galaxies should further
reveal how well the dust properties may be associated with other
characteristics of the ISM, such as the gas density, metallicity,
and/or galaxy type.  In these forthcoming analyses, however, we need
to ensure that the model assumptions are fully considered in the
interpretations.  For instance, in this work we model the dust
emission with single-component SEDs.  As we are focusing on the dust
and gas properties on large scales (100 pc), there is most certainly
temperature variations along the LoS.  Indeed, recent investigations
on SED modelling on smaller scales have suggested that models with
multiple dust components can better reproduce the observed IR emission
\citep[either with varying \Td, \betaeff, or both, see
e.g.][]{Galliano+11, Gordon+14}.  Future efforts including a
consideration of temperature variations may provide additional
insights into the large scale ISM.

\section{Summary} \label{sumsec}

We have introduced a hierarchical Bayesian (HB) method to analyze FIR
and CO observations.  The HB method estimates parameters of the dust
SED; using estimates of the dust surface density \Sigd, a single
temperature \Td\ and spectral index \betaeff, the method calculates
the integrated FIR intensity \Sir\ at each pixel.  Furthermore, the
method simultaneously estimates the linear regression parameters of
the $\log$ \Sir\ $-$ $\log$ \Ico\ relationship.  When assuming that
\Sir\ and \Ico\ faithfully trace the star formation rate \sigsfr\ and
molecular gas surface density \sigmol, respectively, the slope of this
relationship is the Kennicutt-Schmidt index.

We test the HB method on synthetic datasets (Section \ref{testsec}).
Even when the distributions of some key latent parameters are not
normal, which is assumed in the HB model, the range in estimated
parameters include the true underlying values.  We also compare the HB
results with common non-hierarchical techniques.  Due to the
degeneracy between \Td\ and \betaeff\ in the SED, the \chisq\ fit
produces a \Td $-$ \betaeff\ distribution that is biased towards an
anti-correlation.  The HB method explicitly treats the correlation
between \Td\ and \betaeff\ among all pixels, and we demonstrate the
posterior accurately recovers \rhoTB.  Moreover, a \chisq\ fit to the
relationship between \Ico\ and \Sir, or any individual IR intensity,
is biased towards smaller slopes.  This occurs in part because the
\chisq\ analysis overestimates \Sir\ at low densities.

We apply the HB fit to {\it Herschel} IR and {\it NANTEN} CO maps of
the LMC and SMC at 100 pc scales.  The main results of this first
application of the HB methods are as follows:

1) We find a stronger negative correlation between \Td\ and \betaeff\
in the LMC, with \rhoTB $\approx$ $-0.74$, compared to the SMC, with
\rhoTB $\approx -0.15$ .  These results reflect fundamental difference
in the properties of dust and the structure of the ISM between the
Magellanic Clouds.

2) The slopes of the FIR $-$ CO relationship for both galaxies are
similar, falling in the range 1.1 $-$ 1.7.  However, in the SMC the
intercept is nearly 0.4 dex higher.  This difference can be attributed
to the lower metallicity of the SMC.  Due to the paucity of CO, there
is larger \Sir\ per unit \Ico\ in the SMC compared to the LMC, where
the metallicity is about two times larger.  The lower metallicity in
the SMC can also explain the higher overall temperatures and \Sir\ at
a given \Ico.  If there are fixed conversion factors (within a galaxy)
between \Ico\ and \sigmol, and FIR and \sigsfr, then these results
suggest that the Magellanic Clouds have similar Kennicutt-Schmidt
indices.

3) In the LMC, the HB modelling reveals an increase in \Td\ in regions
with the highest CO intensities.  At \Ico\apgt0.5 \Icounits, \Td\
increases from $\approx$20 K to $>$30 K where \Ico $\approx$ 3
\Icounits.  This is indicative of increased dust heating at the
densest regions, likely from newly born stars.  There is also evidence
for increasing \Td\ towards lower gas densities at \Ico\aplt0.1
\Icounits, due to the waning influence of self-shielding in diffuse
regions.

We discuss the SED parameters and KS relationship in Section
\ref{discsec}.  Further investigation of the FIR intensity is
necessary to understand the origins of the difference in \rhoTB\
between the two galaxies.  The difference in KS slopes between the
irregular Magellanic Clouds, where $n$ is clearly above unity, and
normal disk galaxies where $n$ is sub-linear, may be due to
metallicity or other global galaxy properties.  Similar hierarchical
modelling of other galaxies will allow for a more direct comparison
between the dust and gas properties of the ISM under diverse galactic
environments.

\section*{Acknowledgments}
The results presented here have made extensive use of the {\tt ggplot}
library in the R statistical package \citep{ggplot}.  We thank
B. Weiner, A. Stutz, K. Gordon, and B. Groves for insightful
discussion about the physics of the ISM, as well as the anonymous
referee for suggestions that improved this paper.  We are also
grateful to B. Kelly for providing guidance on the statistical
analysis of observational datasets.  RS, SH, RSK, and EP acknowledge
support from the Deutsche Forschungsgemeinschaft (DFG) via the SFB 881
(B1 and B2) ``The Milky Way System,'' and the SPP (priority program)
1573.  RSK also acknowledges support from the European Research
Council under the European Community’s Seventh Framework Programme
(FP7/2007-2013) via the ERC Advanced Grant STARLIGHT (project number
339177).  TL and DR are supported in part by NSF grant AST-1312903.
SH acknowledges financial support from DFG programme HO 5475/2-1.

\bibliography{citations}
\bibliographystyle{mn2e}

\label{lastpage}

\end{document}